\documentclass[twocolumn, twocolappendix]{aastex631}
\usepackage{amssymb}
\usepackage{natbib}
\usepackage{multirow}
\usepackage{hyperref}
\usepackage{tablefootnote}
\usepackage{footnote}
\usepackage{soul}

\begin{document}

\defcitealias{2004Pettini}{PP04}
\defcitealias{2010Mannucci}{M10}
\defcitealias{2013Marino}{M13}
\defcitealias{2018Strom}{S18}
\defcitealias{2018Bian}{B18}
\defcitealias{2012Kennicutt}{KE12}
\defcitealias{2014Steidel}{S14}

\title{Exploring the Relationship Between Stellar Mass, Metallicity, and Star Formation Rate at $z\sim2.3$ in KBSS-MOSFIRE}

\correspondingauthor{Nathalie Korhonen Cuestas}
\email{nathaliekorhonencuestas2029@u.northwestern.edu}

\author[0000-0003-2385-9240]{Nathalie A. Korhonen Cuestas}
\affiliation{Department of Physics and Astronomy, Northwestern University, 2145 Sheridan Road, Evanston, IL 60208, USA}
\affiliation{Center for Interdisciplinary Exploration and Research in Astrophysics (CIERA), Northwestern University, 1800 Sherman Avenue, Evanston, IL 60201, USA}

\author[0000-0001-6369-1636]{Allison L. Strom}
\affiliation{Department of Physics and Astronomy, Northwestern University, 2145 Sheridan Road, Evanston, IL 60208, USA}
\affiliation{Center for Interdisciplinary Exploration and Research in Astrophysics (CIERA), Northwestern University, 1800 Sherman Avenue, Evanston, IL 60201, USA}

\author[0000-0001-8367-6265]{Tim B. Miller}
\affiliation{Center for Interdisciplinary Exploration and Research in Astrophysics (CIERA), Northwestern University, 1800 Sherman Avenue, Evanston, IL 60201, USA}

\author[0000-0002-4834-7260]{Charles C. Steidel}
\affiliation{Cahill Center for Astronomy and Astrophysics, California Institute of Technology, MS 249-17, Pasadena, CA 91125, USA}

\author[0000-0002-6967-7322]{Ryan F. Trainor}
\affiliation{Department of Physics and Astronomy, Franklin \& Marshall College, 637 College Avenue, Lancaster, PA 17603, USA}

\author[0000-0002-8459-5413]{Gwen C. Rudie}
\affiliation{The Observatories of the Carnegie Institution for Sciences, 813 Santa Barbara Street, Pasadena, CA 91101, USA}

\author[0000-0001-5595-757X]{Evan Haze Nuñez}
\affiliation{Cahill Center for Astronomy and Astrophysics, California Institute of Technology, MS 249-17, Pasadena, CA 91125, USA}







\begin{abstract} The metal enrichment of a galaxy is determined by the cycle of baryons in outflows, inflows, and star formation. The relative contribution and timescale of each process sets the relationship between stellar mass, metallicity, and the star formation rate (SFR). In the local universe, galaxies evolve in an equilibrium state where the timescales on which SFR and metallicity vary are comparable, and they define a surface in mass-metallicity-SFR space known as the Fundamental Metallicity Relation (FMR). However, high-redshift observations suggest that this state of equilibrium may not persist throughout cosmic time. Using galaxies from the Keck Baryonic Structure Survey (KBSS) observed with MOSFIRE, we explore the relationship between stellar mass, gas-phase oxygen abundance, and SFR at $z\sim2.3$. Across multiple strong-line calibrations and SFR calculation methods, KBSS galaxies are inconsistent with the locally-defined FMR. We use both parametric and non-parametric methods of exploring a mass-metallicity-SFR relation. When using a parametric approach, we find no significant reduction mass-metallicity relation scatter when folding in SFR as a third parameter, although a non-parametric approach reveals that there could be a weak, redshift-dependent anticorrelation between residual gas-phase oxygen abundance and SFR. Injection-recovery tests show that a significant reduction in scatter requires a stronger anticorrelation between SFR and residual metallicity. Our results suggest that the local FMR may not persist to $z\sim2.3$, implying that $z\sim2.3$ galaxies at this redshift may not be in the equilibrium state described by the FMR and are more similar to higher redshift galaxies. 


\end{abstract}

\keywords{}


\section{Introduction} \label{sec:intro}
\par The flow of gas into and out of a galaxy sets its evolutionary trajectory. Metal-poor gas can be accreted from the circumgalactic medium or through interactions with nearby lower-mass galaxies, creating a gas-rich, metal-poor environment and triggering episodes of star formation. Eventually, these same stars will enrich the interstellar medium (ISM) through stellar winds and supernova explosions. Galactic winds driven by supernovae, massive stars, and active galactic nuclei (AGN) can transport gas out of the galaxy, depleting the gas reservoir, curtailing star formation, and transporting metals out of the galaxy. Understanding how these processes influence the galactic ecosystem and change over cosmic time is necessary to gain a full picture of galaxy growth and evolution.

\par Elemental abundances are sensitive to the relative contributions of metal-poor inflows, gas-rich outflows, and star formation. As a result, chemical scaling relations are a key probe of the baryon cycle. \cite{1979Lequeux} first identified a correlation between the luminosity, mass, and metallicity of H \textsc{ii} regions in irregular and blue compact galaxies. A similar luminosity-metallicity relation was observed by \cite{2002Garnett} in irregular galaxies and by \cite{2004Pilyugin} in spiral galaxies. Using a sample of $\sim$53,000 Sloan Digital Sky Survey \citep[SDSS;][]{2000York} galaxy spectra, \cite{2004Tremonti} robustly characterized the mass-metallicity relation (MZR). \cite{2004Tremonti} found that lower-mass galaxies are more metal-poor than higher-mass galaxies, and the slope of the relationship becomes shallower at higher masses. The slope of the MZR can only be reproduced by models that consider outflows and inflows \citep{2007Kobayashi, 2008Finlator, 2017Dave}, demonstrating that galaxies do not evolve as ``closed-box" systems \citep{2004Tremonti, 2008Finlator}. 
 
\par Since the analysis by \cite{2004Tremonti}, the MZR has been observed at a range of redshifts 
\citep[$z\lesssim1$:][$z\sim2$: \citealt{2006Erb_a, 2008Panter, 2009Mannucci, 2014Steidel, 2018Sanders}; \citealt{2022Strom}, and $z\gtrsim3$: \citealt{2008Maiolino, 2014Troncoso, 2016Onodera, 2021Sanders, 2023Curti, 2023Nakajima}, among many others]{2005Savaglio, 2008Liu, 2008Cowie, 2009Perez-Montero}. The MZR evolves towards lower normalizations at higher redshifts, meaning that at a fixed stellar mass, galaxies are more metal-poor at higher redshifts. A variety of processes could drive the MZR's evolution towards lower metallicities. Higher gas fractions, stronger outflows due to either enhanced star formation or AGN activity, and higher inflow rates of metal-poor gas at higher redshifts could all lead to the observed redshift evolution \citep[e.g.,][]{2005Savaglio, 2006Erb_a, 2008Maiolino, 2009Mannucci}. 

\par The dispersion of galaxies about the MZR ($\sim0.1$ dex in the \citealt{2004Tremonti} sample) has also been extensively studied. Offsets relative to the MZR have been correlated with star formation rate (SFR), specific star formation rate (sSFR), gas mass, galaxy size, and galaxy color \citep{2004Tremonti, 2008Ellison, 2010Mannucci, 2014Salim, 2016Bothwell}. For local galaxies at fixed mass, metallicity is inversely correlated with SFR, leading both \cite{2010Lara-Lopez} and \cite{2010Mannucci} to suggest the presence of a more fundamental, three-parameter relationship linking stellar mass ($M_\star$), gas-phase metallicity ($Z_g$), and SFR that underpins the MZR and can reproduce both the MZR's observed scatter and its redshift evolution. 
\cite{2010Mannucci} specifically defined a three-dimensional surface in $M_\star$-$Z_g$-SFR space, which local galaxies populate with very little scatter ($\sim0.05$ dex for SDSS galaxies). Importantly, this Fundamental Metallicity Relation (FMR) was hypothesized to be redshift-invariant, implying a mechanism for galaxy growth and enrichment that is also redshift invariant. 
\par The FMR is consistent with the gas-regulator model described by \cite{2013Lilly}, assuming constant star formation efficiency and outflow mass loading factors. In this model, galaxies are continuously fed by a slowly-evolving infall of gas, the SFR is instantaneously regulated by the gas mass of the galaxy, and mass loss rates scale with SFR, thereby resulting a redshift-invariant FMR. 

\par The presence (or absence) of the FMR at high redshift has been widely investigated \cite[e.g.,][]{2014Troncoso, 2014ZahidKashino, 2015Maier, 2015Yabe, 2016Kacprzak, 2016Onodera, 2021Sanders}, and the James Webb Space Telescope (JWST) has allowed FMR studies to push to ever higher redshifts \citep[e.g.,][]{2023Curti, 2023Nakajima, 2024Curti}. However, these works vary significantly in the methods used to measure oxygen abundance and assess consistency with the local FMR. Oxygen abundance is often estimated using strong-line calibrations \citep[e.g.,][]{2014Steidel, 2014ZahidKashino, 2016Wuyts, 2017Kashino, 2024Curti}, although sometimes direct measurements via the $T_e$ method are accessible, either in individual spectra \citep{2017Calabro, 2023Curti, 2023Nakajima, 2024Revalski} or in composite spectra \citep{2021Sanders, 2021Topping}. Photoionization models are also sometimes used \citep{2014Maier, 2017Calabro}. Galaxies may also be treated individually \citep{2023Curti} or binned by mass, SFR, or redshift \citep[e.g.,][]{2015Sanders, 2015Yabe, 2018Sanders}. 
\par The FMR and more general $M_\star$-$Z_g$-SFR relationships are probed in three main ways: using a similar methodology to \cite{2010Mannucci} to find the projection of least scatter \citep[e.g.,][]{2013Andrews, 2016Onodera, 2021Sanders}, direct comparison to a locally-defined FMR \citep[e.g.,][]{2014Maier, 2015Sanders, 2015Yabe, 2023Curti, 2024Curti}, or by quantifying the correlation between residual metallicity (defined relative to the median for galaxies of the same mass) and SFR \citep[e.g.,][]{2014Salim, 2014ZahidKashino, 2016Wuyts}. 
\par Given the broad range of possible approaches, consensus remains elusive. Some find high-redshift galaxies to be consistent with a locally-calibrated FMR plane \citep{2015Maier, 2015Yabe, 2016Kacprzak, 2021Topping}, supporting the existence of a redshift-invariant FMR, while others observe a $M_\star$-SFR-$Z_g$ relationship but find their galaxies to be offset from local FMR planes \citep{2014ZahidKashino, 2015Salim, 2017Kashino, 2018Sanders}, instead suggesting that the FMR evolves with redshift. Still others find no evidence of the FMR \emph{or} a three parameter correlation \citep{2014Steidel, 2016Wuyts}. 

\par This paper aims to characterize the relationship between stellar mass, gas-phase oxygen abundance, and SFR at $\langle z\rangle = 2.29$ using a large, homogeneously selected sample of galaxies from the complete Keck Baryonic Structure Survey (KBSS) and determine the extent to which the data are consistent with the locally-defined FMR. By employing different analysis methods we also aim to explore how our results, and the results of other studies, depend on which method is used to define and characterize the FMR. For the sake of clarity, we will use FMR to specifically refer to a redshift-invariant surface in $M_\star$-$Z_g$-SFR space, as proposed by \cite{2010Mannucci}. Any other correlations between $M_\star$, $Z_g$, and SFR will be referred to as a $M_\star$-$Z_g$-SFR relationship. The galaxy sample and measured properties are introduced in Section \ref{sec:kbss}. The star forming main sequence (SFMS) and MZR for KBSS galaxies are reported in Sections \ref{sec:sfms} and \ref{sec:mzr}. In Section \ref{sec:am13_comp} we compare KBSS galaxies to the locally-defined FMR, and in Sections \ref{sec:alpha_method} and \ref{sec:nonparam}, we explore parametric and non-parametric approaches. In Section \ref{sec:discussion} we attempt to reconcile the results of both parametric and non-parametric methods (Section \ref{sec:injectionrecovery}), compare our results to both local and high-$z$ surveys (Section \ref{sec:othersamples}), and discuss potential physical implications (Section \ref{sec:physimplications}). Lastly, we present our conclusions in Section \ref{sec:conclusions}.

\par Throughout this paper, we assume a $\Lambda$CDM cosmology with $H_0=70\, \mathrm{km\,s^{-1}\, Mpc^{-1}}, \, \Omega_\Lambda=0.7$, and $\Omega_{m}=0.3$ and adopt the Solar oxygen abundance value reported in \cite{2009Asplund}, $12+\log(\mathrm{O/H})_\odot=8.69$, but assume $Z_\odot = 0.02$ as this is the value adopted by BPASS. Individual emission lines are referred to using their vacuum wavelength in Angstroms. 

\section{The Keck Baryonic Structure Survey} \label{sec:kbss}
\par
\begin{figure}
    \centering
    \includegraphics[width=1.0\linewidth]{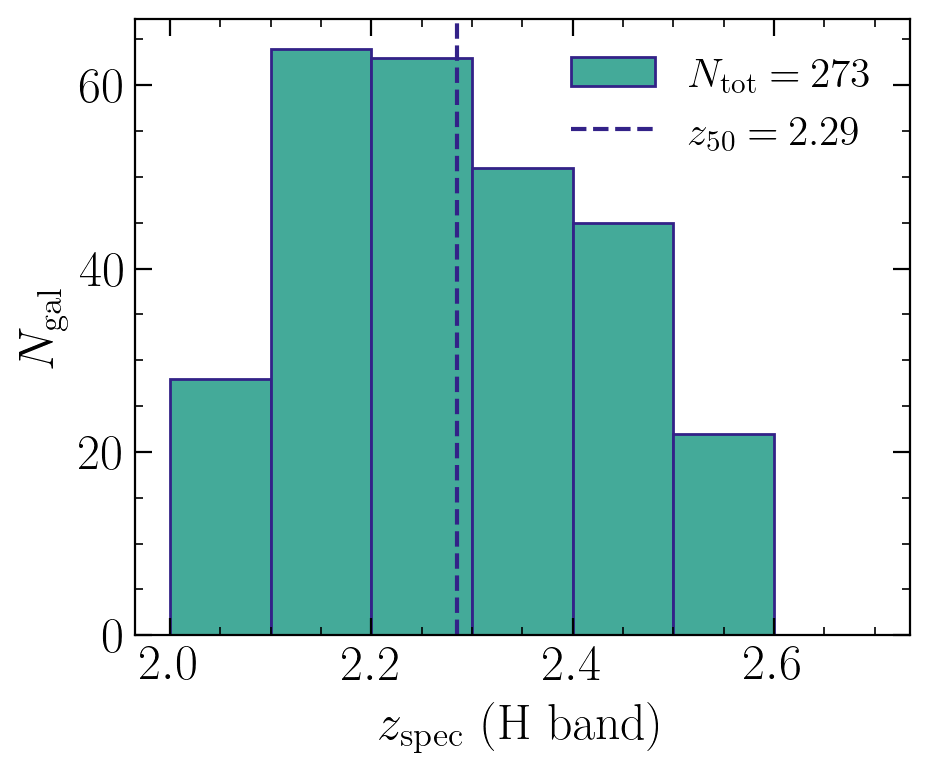}
    \caption{The redshift distribution of the O3N2-selected KBSS galaxies. The sample probes a redshift range of $2.0<z<2.6$, with a median $z_{50}=2.29$.}
    \label{fig:zhist}
\end{figure}

 The galaxies analyzed in this paper were observed as part of the Keck Baryonic Structure Survey \citep[KBSS;][]{2012Rudie, 2014Steidel}. KBSS is a targeted spectroscopic survey designed to observe galaxies at $z\sim1.5-3.5$ (see Figure \ref{fig:zhist} for the redshift distribution of galaxies analyzed in this paper) in fifteen quasar fields over a total area of $\sim0.24\,\mathrm{deg}^2$, with substantial rest-UV and rest-optical imaging and spectroscopy conducted using the Keck LRIS \citep{1995Oke} and MOSFIRE \citep{2012McLean} instruments. Here, we only consider galaxies with rest-optical spectra, which make up the KBSS-MOSFIRE subsample. Importantly, the strong rest-optical lines produced in H \textsc{ii} regions are redshifted into the NIR \emph{J}, \emph{H}, and \emph{K} bands at $1.9<z<2.7$, with a subset of these strong lines accessible at $z\sim1.5$ and $z\sim3$. Most KBSS galaxies were selected based on their rest-UV colors \citep[for further detail, see][]{2004Adelberger, 2004Steidel}. Some additional objects are selected based on their $\mathcal{R}$-K colors in order to mitigate a slight preference towards highly star-forming galaxies \citep{2017Strom}. 
\par For further detail on the KBSS survey design, photometric data, spectroscopy, and data reduction, see \cite{2004Steidel, 2012Reddy, 2014Steidel} and \cite{2017Strom}. Here, we discuss some of the more pertinent aspects of KBSS-MOSFIRE. 

\subsection{Stellar Mass}\label{sec:mass}
\par To estimate the stellar mass ($M_\star$) of each galaxy and obtain object-by-object posteriors, we use the spectral energy distribution (SED)-fitting code \textsc{Bagpipes} \citep{2018Carnall} and broadband photometry. All fifteen KBSS quasar fields are observed in at least eight photometric bands including the optical $U_n$, $G$, and $\mathcal{R}$s bands, near-IR $J$ and $K_s$, and mid-IR IRAC Ch1 and 2. Fourteen fields have photometry in WFC3 F140W and F160W (Q1549 lacks F140W coverage and Q0821 lacks F160W coverage). Nine of the fields are also observed in the near-IR \emph{H} band with Magellan/FourStar and six fields are observed in IRAC Ch 3. Photometry in B band, WFC3 F110W, F125W, and F814W is also available for select fields. For further detail on the photometry in KBSS fields, see \cite{2019Theios}. The photometry has also been corrected for the contribution from emission lines. 
\par The \textsc{Bagpipes} model galaxy spectra use \textsc{BPASS} \citep[v2.2.1][]{2018Stanway} binary stellar population models with a \cite{2001Kroupa}-like IMF with a upper mass slope of $-2.35$ and a maximum stellar mass of $100 M_\odot$. We do, however, rescale the resulting stellar masses to a \cite{2003Chabrier} IMF to allow for closer comparison to other samples. The nebular continuum is modeled using \textsc{Cloudy} \citep{2017Ferland}, where the input spectrum is taken from the \textsc{BPASS} grids \citep{2018Carnall}. Notably, we do not consider the effect of nebular emission lines on the model galaxy SED, since the photometry has already been corrected for the contribution of emission lines. We adopt the SMC \citep{2003Gordon} dust reddening curve, which \cite{2019Theios} found to yield the best agreement between H$\alpha$-based SFRs and SED-based SFRs when applied to the stellar continuum of KBSS galaxies. We limit the metallicity range to $[0, 0.5]Z_\odot$, assuming the same value for both the stellar and gas-phase metallicities; the log of the ionisation parameter (defined as the ratio between ionizing photon density and hydrogen number density $U\equiv n_{\gamma, i}/n_H$) range to $[-3.5, -2.5]$; the dust attenuation $A_V$ range to $[0, 2]$ magnitudes; and the log of the stellar mass range to $[0, 15]$. The redshift of a galaxy was held fixed at the spectroscopic redshift. 
\par We implement a non-parametric, continuity star formation history (SFH), which has been shown to more accurately recover galaxy properties and their associated uncertainties and introduce less bias than a parametric star formation history \citep{2019Carnall, 2019Leja}. To fit the SFH, \textsc{Bagpipes} fits the change in SFR ($\Delta\log(\mathrm{SFR})$) between set time bins. We set the bins to be 0, 10, 100, 250, 500, 1000, 1500, 2000, 2500, 3000, and 3500 Myr before the time of observation. In practice, the oldest time bin is sometimes not considered as it is excluded by the galaxy's spectroscopic redshift. The prior on $\Delta\log(\mathrm{SFR})$ is set as a Student's-t distribution with a mean value of 0, a scale factor of 0.3, and 2 degrees of freedom, as in \cite{2019Leja}, which has a preference towards smaller changes in SFR between time bins, resulting in generally smoother SFHs. 
\par The results of our SED-fitting are broadly consistent with previously published KBSS stellar masses \citep{2019Theios}. The primary difference is the adoption of a non-parametric SFH; previously, a constant SFH had been assumed. A non-parametric SFH tends to produce slightly higher $M_\star$ estimates as it allows for a more extended SFH, during which galaxies build up more mass. The difference is small ($\sim0.06$ dex, or a $14\%$ increase in $M_\star$): the median stellar mass assuming a constant SFH is $M_\star\sim 5.6\times 10^9 M_\odot$ and the median stellar mass assuming a non-parametric SFH is $M_\star\sim6.4\times10^9 M_\odot$. Further comparison between the stellar masses used here and the stellar masses reported in \cite{2019Theios} will be discussed in Strom et al. (in prep). 

\subsection{Strong Line Measurements and Dust Correction}
\par Emission line measurements were made from the extracted 1D MOSFIRE spectra \citep{2014Steidel}. The continuum is estimated using stellar population synthesis models from \cite{2003Bruzual}
fit to the observed galaxy SED, which self-consistently account for the effects of dust extinction and stellar Balmer absorption features \citep[for further detail see][]{2017Strom}. 
Strong lines in a single MOSFIRE band are fit using Gaussians with a single redshift and line width. The flux ratios of the [O~\textsc{iii}]$\lambda\lambda4959,5008$ and [N \textsc{ii}]$\lambda\lambda6549,6585$ doublets were fixed at 1:3, as set by atomic physics \citep{2000Storey}. Line fluxes were corrected for potential slit losses using the method detailed in \cite{2017Strom}. In the case of some higher-$z$ galaxies, [O~\textsc{iii}]$\lambda 5007$ falls outside of \emph{H} band, and so cannot be directly measured in the spectrum. In these cases, we used the detection of [O~\textsc{iii}]$\lambda 4959$ to estimate the flux of [O \textsc{iii}]$\lambda 5007$ using the same 1:3 flux ratio. 

\par To account for the effect of dust attenuation toward star-forming regions, we adopt a Milky Way dust extinction curve with $R_V=3.1$ \citep{1989Cardelli}, consistent with previous KBSS studies \citep{2019Theios}. Extinction is estimated using the Balmer decrement, assuming a theoretical value of $\mathrm{H}\alpha/\mathrm{H}\beta=2.86$ \citep[Case B;][]{1989Osterbrock}. We imposed a selection criterion that the SNR of $\mathrm{H}\alpha/\mathrm{H}\beta>5$. Given that the uncertainty of extinction corrections scale as $\sim(\mathrm{H}\alpha/\mathrm{H}\beta)^{k_\lambda}$, where $k_\lambda$ is the value of the reddening curve at a specific wavelength (for reference, $k_\mathrm{H\alpha}=2.53$), small uncertainties in the Balmer decrement can propagate as large uncertainties in the corrected line flux. In cases where the Balmer decrement was found to be $<2.86$, zero extinction was assumed. Dust-corrected line fluxes were only used in SFR calculations and not in the calculation of strong-line ratios. The strong-line ratios used to infer oxygen abundance in this paper require emission lines that are fairly close to each other in wavelength, meaning that the ratio is only weakly affected by dust attenuation and a reddening correction is not necessary.

\subsection{Gas-Phase Oxygen Abundance}\label{sec:oxygen}
The most accurate measurements of gas-phase oxygen abundances require an electron temperature ($T_e$) measurement. The auroral emission lines needed measure $T_e$ are intrinsically much fainter than strong nebular emission lines. For example, the auroral line [O \textsc{iii}]$\lambda4363$ can be $\sim100\times$ fainter than the strong nebular line [O \textsc{iii}]$\lambda5007$, making them difficult to detect at high redshift \citep[although this is changing in the era of JWST, see for example][]{2022Schaerer, 2023Curti, 2024Hsiao, 2024Rogers}. Alternatively, we can use strong-line calibrations, which correlate the ratio of strong rest-optical nebular emission lines with oxygen abundance. The calibration can be based on samples with direct $T_e$ measurements \citep[e.g.,][]{2013Marino, 2018Bian}, photoionization modeling \citep[e.g.,][]{1991McGaugh, 2018Strom}, or on a combination of the two \citep[e.g.,][]{2004Pettini}. 
\par There are several limitations to the use of strong-line calibrations. Many calibrations are based on low-$z$ galaxies, with systematically different ISM abundances and ionizing radiation \citep{2014Masters, 2014Steidel, 2015Shapley, 2017Strom, 2024Shapley_a}, meaning that applying local calibrations at high redshift can introduce bias. Using calibrations based on H \textsc{ii} regions or low-redshift analogues to high-redshift galaxies can go some way to addressing this issue, as the calibration sample may reflect ISM conditions more similar to those in high-redshift galaxies. Improved observational capabilities due to JWST has also allowed ongoing work to recalibrate these strong-line calibrations \emph{in-situ} by directly observing auroral lines at ever higher redshifts \citep{2024Laseter, 2024Sanders, 2025Scholte}. 
\par However, strong-line ratios are used throughout the literature to produce MZRs \citep[e.g.,][]{2005Savaglio, 2006Erb_a, 2024Revalski, 2024Curti} and establish the existence of the FMR \citep[e.g.,][]{2010Mannucci, 2015Maier, 2020Curti}, so with their limitations in mind, we apply strong-line calibrations to KBSS galaxies to facilitate comparison with previous work. 
We test two different strong-line ratios, O3N2 and N2, which we define as 
\begin{equation}
    \mathrm{O3N2} \equiv \log\left(\frac{\mathrm{[O \,\textsc{iii}]}\lambda 5008}{\mathrm{H}\beta}\right) - \log\left(\frac{\mathrm{[N \,\textsc{ii}]}\lambda 6585}{\mathrm{H}\alpha}\right)
\end{equation}
\begin{equation}
    \mathrm{N2} \equiv \log\left(\frac{\mathrm{[N \, \textsc{ii}]}\lambda 6585}{\mathrm{H}\alpha}\right)
\end{equation}
\par O3N2 and N2 are both sensitive to ionization parameter as well as oxygen abundance \citep{2002Kewley}, introducing significant systematic uncertainty to individual oxygen abundance measurements. 
Another widely-used strong-line indicator of oxygen abundance is R23$\equiv \log(([\mathrm{O \,III}]\lambda\lambda4960,5008 + [\mathrm{O II}]\lambda\lambda3727,3729)/\mathrm{H}\beta)$, which is less sensitive to ionization parameter than O3N2 and N2. Additionally, the R23 ratio can be modified by weighing $[\mathrm{O \,III}]\lambda\lambda4960,5008$ and $[\mathrm{O II}]\lambda\lambda3727,3729$ such that the dependence on ionization parameter is minimized \citep[][]{2024Laseter, 2025Scholte}. However, R23 is double-valued and sensitive to stellar iron abundance at its turning point \citep[e.g.,][]{2018Strom}. Since many KBSS galaxies lie near the turning point and some lie beyond the range of R23 allowed by most strong-line calibrations, R23-based oxygen abundances are not investigated in detail here. However, it is worth noting that we have conducted our analysis using galaxies with reliably-measured R23 and the R23 calibration from \cite{2018Strom} and find the same qualitative results as what is reported in this paper.
\par For each line ratio, we test four different strong-line calibrations. Each calibration differs in terms of the parent sample and method used to measure abundance. Moreover, each calibration produces an MZR with a different slope, normalization, and intrinsic scatter. Comparing different calibrations therefore allows us to test whether the detection of the FMR differs depending on the associated MZR slope, MZR scatter, or chosen strong-line ratio. 
\par The calibrations all take the form 
\begin{equation}\label{calibration}
    12+\log(\mathrm{O/H}) = A - B\times \mathrm{R}
\end{equation}
Where $\mathrm{R}$ is the line ratio, and $A$ and $B$ are constants unique to the specific calibration. For a summary of the different calibrations we use, see Table \ref{tab:calibrations}. The \citet[][hereafter \citetalias{2004Pettini}]{2004Pettini} calibration is based on observations of extragalactic H \textsc{ii} regions where either $T_e$ could be directly measured or the spectrum could be robustly reproduced by photoionization models. The calibration has been applied to different samples at comparable redshifts to KBSS (e.g., \citealt{2021Topping}), allowing for a direct comparison between our results. \citet[][hereafter \citetalias{2014Steidel}]{2014Steidel} use the same observations as \citetalias{2004Pettini}, but base the calibration only on H \textsc{ii} regions where oxygen abundance was measured via the $T_e$ method and the value of N2 was within the range observed in KBSS-MOSFIRE galaxies. \citet[][hereafter \citetalias{2013Marino}]{2013Marino} also use $T_e$ measurements of H \textsc{ii} regions, which were observed in the CALIFA survey \citep{2012Sanchez, 2013Sanchez}. The direct metallicity measurement is combined with the metallicity estimated from strong nebular oxygen, nitrogen, and sulfur emission lines, as described in \cite{2010Pilyugin} with the goal of producing a more accurate overall calibration. The remaining calibrations are all based on observations of galaxies rather than H \textsc{ii} regions. \citet[][hereafter \citetalias{2018Bian}]{2018Bian} identify low-redshift galaxies that lie on the $z\sim2$ star forming sequence in the N2-BPT diagram as potential analogs to high-redshift galaxies. The calibration is based on direct measurements of $T_e$ in these galaxies. Lastly, the calibrations from \citet[][hereafter \citetalias{2018Strom}]{2018Strom} are based on 150 KBSS galaxies where the abundance was estimated using photoionization models. Therefore, we are able to compare calibrations which use $T_e$ measurements, photoionization models, or a combination and are based on populations of local H~\textsc{ii} regions, local analogues to high-$z$ galaxies, and $z\sim2.3$ galaxies.  
\par To ensure that the necessary lines are well-measured, we select galaxies where the SNR of all emission lines used in a given strong-line calibration is $>3$ and the SNR of H$\alpha$/H$\beta > 5$. This criteria results in a sample of 273 galaxies with well-measured O3N2 and 274 galaxies with well-measured N2.
\begin{table}[]
    \centering
    \caption{Summary of the eight different line ratios used in this paper. $A$ and $B$ refer to the constants in Equation \ref{calibration}.}
    \begin{tabular}{c| c c c}
    \hline\hline
         Calibration & Line Ratio & $A$ & $B$ \\ \hline\hline
         \cite{2004Pettini} & O3N2 & 8.73 & 0.32 \\
         \cite{2013Marino} & O3N2 & 8.533 & 0.214 \\
         \cite{2018Bian} & O3N2 & 8.97 & 0.39 \\ 
         \cite{2018Strom} & O3N2 & 8.75 & 0.21 \\\hline
         \cite{2013Marino} & N2 & 8.743 & $-0.462$ \\
         \cite{2014Steidel} & N2 & 8.62 & $-0.36$ \\
         \cite{2018Bian} & N2 & 8.82 & -0.49 \\ 
         \cite{2018Strom} & N2 & 8.77 & $-0.34$ \\\hline\hline
    \end{tabular}
    \label{tab:calibrations}
\end{table}

\subsection{Star-Formation Rate}\label{sec:sfrs}
\par SFR can be estimated as part of the SED fitting, but due to the degeneracy with fit parameters such as age and stellar mass, we prefer SFR estimates derived from the H$\alpha$ luminosity ($\mathcal{L}_{\mathrm{H}\alpha}$). $\mathcal{L}_{\mathrm{H}\alpha}$ must be multiplied by a conversion factor to determine SFR \citep[e.g.,][hereafter \citetalias{2012Kennicutt}]{2012Kennicutt}. However, the conversion factor is typically based on local stellar populations and ISM conditions, and may not be reflective of the ionizing photon production at high redshift. Galaxies at higher redshifts have systematically lower gas-phase metallicities, meaning that young, massive stars that produce much of the Lyman continuum are also more metal-poor. Metal-poor stars produce more ionizing photons per unit mass than metal-rich stars, and as a result, there are more photons available for absorption and re-emission in the ISM. Therefore, a smaller number of metal-poor stars can produce the same $\mathcal{L}_{\mathrm{H}\alpha}$ as a larger population of more metal-rich stars \citep{2013Zhang}. Due to the comparatively low contribution of Type Ia supernova enrichment, the abundance \emph{patterns} in high-$z$ galaxies also differ significantly from solar: high-$z$ galaxies tend to show an enhanced $\alpha$-element abundance compared to the iron abundance \citep{2020Kobayshi}. Consequently, high-$z$ galaxies with sub-solar gas-phase oxygen abundances will have even more sub-solar [Fe/H]. Assuming that [Fe/H] traces stellar metallicity, this corresponds to $Z_\star/Z_\odot\sim0.1$ for $z\sim2-3$ galaxies \citep{2016Steidel, 2018Strom, 2020Topping}.
Neglecting the relationship between metallicity and ionizing photon production can lead to an overestimation of the SFR at high redshift. 
\par To capture the impact of metallicity on the ionizing spectrum, we adopt a metallicity-dependent conversion factor between $\mathcal{L}_{\mathrm{H}\alpha}$ and SFR. This relationship takes the form
\begin{equation}\label{sfr}
    \log(\mathrm{SFR}) = \log(\mathcal{L}_\mathrm{H\alpha}) - \log(C(Z_\star))
\end{equation}
Where $\mathcal{L}_\mathrm{H\alpha}$ is the H$\alpha$ luminosity in erg s$^{-1}$, SFR is given in $M_\odot \mathrm{yr}^{-1}$, and $C(Z_\star)$ is the metallicity-dependent conversion factor. 
\begin{figure}
    \centering
    \includegraphics[width=1.0\linewidth]{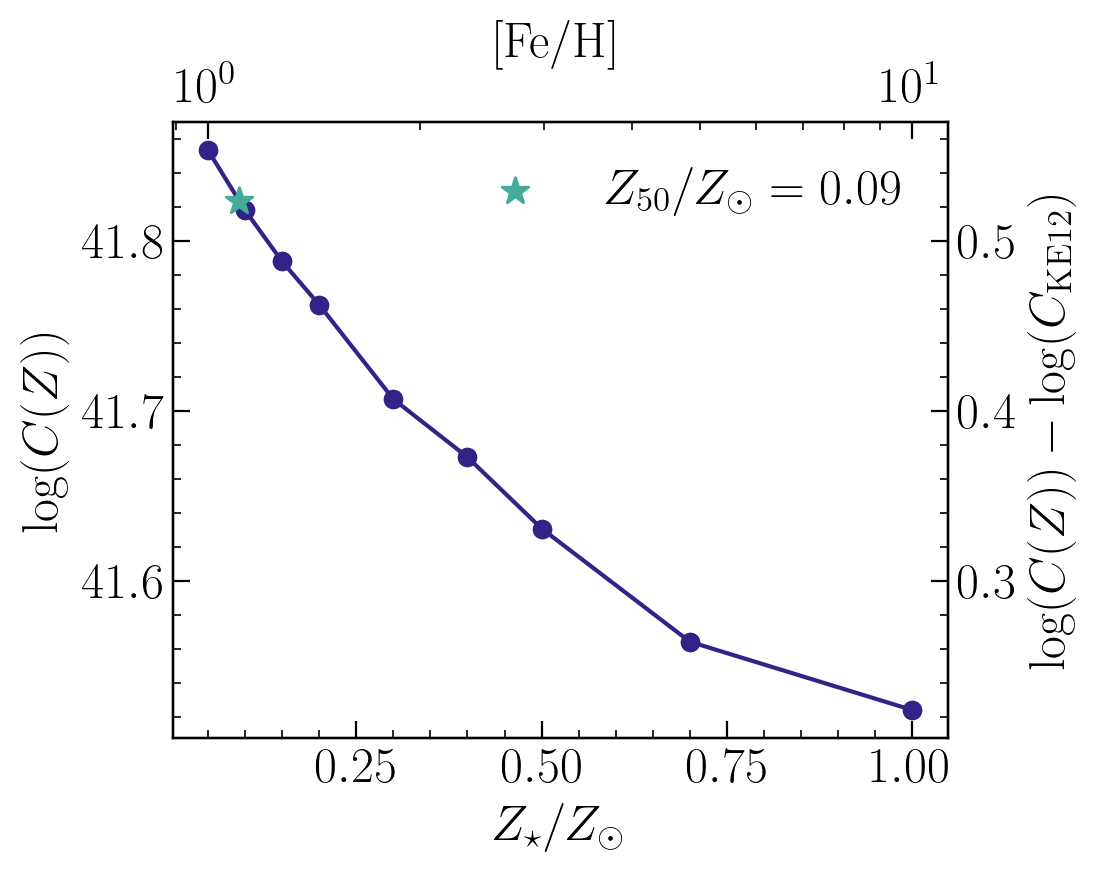}
    \caption{$\mathcal{L}_\mathrm{H\alpha}$ to SFR convsersion factor vs. stellar metallicity. Values calculated from BPASS models are shown in purple, and the median stellar metallicity in the KBSS sample ($Z_{50}/Z_\odot = 0.09$) is shown by the teal star. Adopting $\log(C(Z_\star))$ results in a lower SFR than what is predicted by \citetalias{2012Kennicutt} since $\log(C(Z_\star))$ remains larger than $\log(C_\mathrm{KE12})=41.30$ at all $Z_\star\leq Z_\odot$, and the discrepancy is larger for more metal-poor galaxies. }
    \label{fig:sfr_conversion}
\end{figure}
$C(Z_\star)$ is calculated using model SEDs at different stellar metallicities produced using BPASS \citep[v2.2.1][]{2018Stanway}, assuming a constant SFH over 100 Myr, a \cite{2003Chabrier} IMF with a maximum stellar mass of $100\,M_\odot$, and including the effects of binary stars. Using $f_\lambda$ from the BPASS SED, we integrate up to $912 \,\mathrm{\AA}$ to calculate the ionizing photon production:
\begin{equation}
    N(H^0)=\int_0^{912\,\mathrm{\AA}} \frac{\lambda f_\lambda}{hc} d\lambda
\end{equation}
Then, using Equation 9 in \cite{1995Leitherer}, we calculate $\mathcal{L}_{\mathrm{H}\alpha}$ per $M_\odot \, \mathrm{yr}^{-1}$ of star formation:
\begin{equation}
    \mathcal{L}_{\mathrm{H}\alpha} \mathrm{[erg \, s^{-1}]} = 1.36\times10^{-12} N(H^0) [\mathrm{s}^{-1}]
\end{equation}
\begin{table}[]
    \centering
    \caption{Metallicity-dependent H$\alpha$ luminosity to SFR conversion factors after 100 Myr of constant star formation.}
    \begin{tabular}{c|c}
    \hline\hline
      $Z_\star$ & $\log(C(Z_\star))$ \\ \hline\hline
        0.001 & 41.680 \\
        0.002 & 41.647 \\
        0.003 & 41.619 \\
        0.004 & 41.595 \\
        0.006 & 41.544 \\
        0.008 & 41.512 \\
        0.010 & 41.473 \\
        0.014 & 41.411 \\
        0.020 & 41.373 \\\hline\hline
    \end{tabular}
    \label{tab:sfr_conversions}
\end{table}
The resulting conversion factors are listed in Table \ref{tab:sfr_conversions} and are shown in Figure \ref{fig:sfr_conversion}, which highlights the importance of considering the effect of metallicity. $C(Z_\star)$ decreases monotonically with $Z_\star$ and is higher than the \citetalias{2012Kennicutt} conversion factor, rescaled to a \cite{2003Chabrier} IMF, at solar metallicity ($\log(C_\mathrm{KE12})=41.30$). At the median metallicity of KBSS galaxies in the O3N2 sample ($Z_{50}=0.09Z_\odot$, teal star in Figure \ref{fig:sfr_conversion}), the \citetalias{2012Kennicutt} conversion factor is 0.35 dex lower than the BPASS conversion factor, leading to a 0.35 dex overestimation of $\log(\mathrm{SFR})$. 
\par For each galaxy we adopt a stellar metallicity based on that galaxy's inferred iron abundance, which is determined from their oxygen abundance and assuming the oxygen and iron MZRs have the same slope, but different normalizations \citep{2019Cullen, 2020Sanders, 2020Topping, 2022Strom, 2024Stanton}. We adopt a constant offset for all galaxies of $[\mathrm{O/Fe}] = 0.35$ \citep{2022Strom} and infer [Fe/H] according to 
\begin{equation}
    [\mathrm{Fe/H}] = \log(\mathrm{O/H}) - \log(\mathrm{O/H})_\odot -0.35
\end{equation}
\par
\begin{figure}
    \centering
    \includegraphics[scale=0.65]{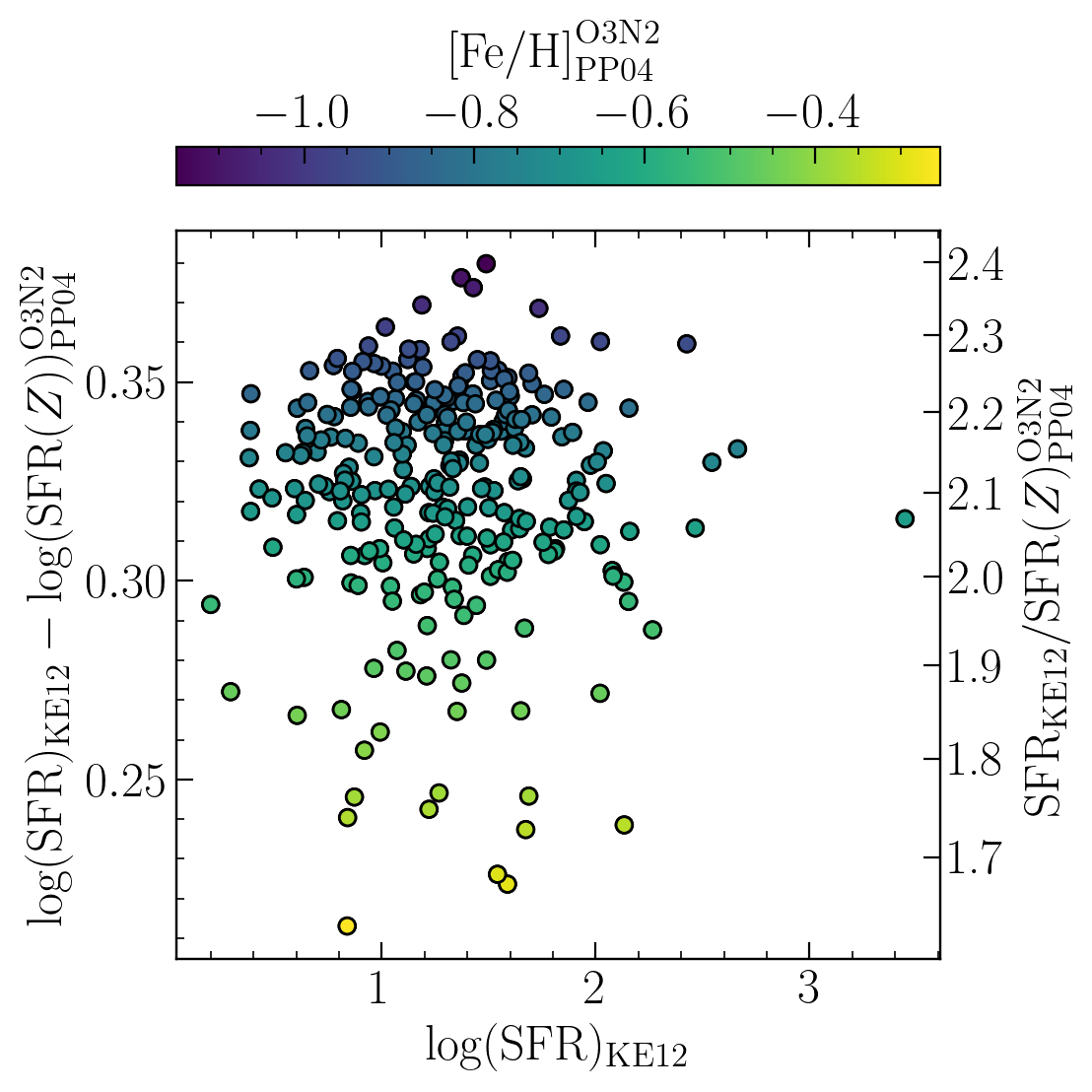}
    \caption{Difference between $\log(\mathrm{SFR_{KE12}})$ and $\log(\mathrm{SFR}(Z))$ vs. $\log(\mathrm{SFR_{KE12}})$, colour-coded by $[\mathrm{Fe/H}]$ for the KBSS O3N2 sample. [Fe/H] is defined as [O/H]$-0.35$ dex, where [O/H] is estimated using the \citetalias{2004Pettini} O3N2 calibration. For all KBSS galaxies, $\mathrm{SFR_{KE12}}>\mathrm{SFR}(Z)$, with the discrepancy being larger for more metal-poor galaxies.}
    \label{fig:sfr_hist}
\end{figure}

 The effect of adopting a metallicity-dependent conversion factor is twofold. The first order effect is a systematic shift towards a lower SFR. This effect arises from the fact that $\log(C(Z_\star)) > \log(C_\mathrm{KE12})$ at all $Z_\star<Z_\odot$ (see Figure \ref{fig:sfr_conversion}). The median $Z_\star$-dependent SFR, using the \citetalias{2004Pettini} calibration ($\mathrm{SFR}(Z)_\mathrm{PP04}^\mathrm{O3N2}$) is $9.3 M_\odot \mathrm{yr}^{-1}$, while the median SFR estimated by the \citetalias{2012Kennicutt} ($\mathrm{SFR_{KE12}}$) relation is $30.9 M_\odot \mathrm{yr}^{-1}$ for galaxies in the O3N2 sample, reflecting the systematic shift towards lower SFR. It should be noted that this bulk shift is not specifically due to the metallicity dependence in the conversion factor: adopting a single conversion factor \citep[e.g., $\log(C)=41.67$ as in][]{2023Shapley_b} that better reflects the photon production of metal-poor stars can create the same effect. The second order effect can be clearly seen in Figure \ref{fig:sfr_hist}: more metal-poor galaxies have a larger discrepancy between $\mathrm{SFR_{KE12}}$ and $\mathrm{SFR}(Z)$. The magnitude of the effect is significantly smaller than the overall shift towards lower SFR, but is nonetheless significant and can only be accounted for using a metallicity-dependent conversion factor. Since stellar metallicity is being inferred from gas-phase oxygen abundance, the magnitude of the effect is dependent on the calibration used to estimate oxygen abundance. Testing different strong line ratios and calibrations allows us to quantify how this inter-calibration variation can affect the characterization of a $M_\star$-$Z_g$-SFR correlation. 
\par Given the crucial role of SFR in the formulation of the FMR, one of the principal aims of this paper is to investigate how a $Z_\star$-dependent SFR may impact the detection of the FMR. Throughout the paper, we conduct all of the analysis using both SFR estimates to allow for direct study of the effect of adopting a metallicity-dependent SFR conversion factor as well as comparison with earlier work. 

\section{The Relationship Between $M_\star$, Metallicity, and SFR} \label{sec:fmr}
\par The combination of SFMS and MZR sets the relationship between $M_\star$, $Z_g$, and SFR. The FMR suggests that the redshift evolution of the SFMS and MZR towards higher SFR and lower $Z_g$ at fixed $M_\star$ respectively, can be captured by a three-dimensional, redshift-invariant surface in $M_\star$-$Z_g$-SFR space. The surface is expected to be roughly planar \citep{2010Mannucci} in the stellar mass and SFR regime probed by KBSS. In the following sections we present two limiting projections of $M_\star$-$Z_g$-SFR space: the SFMS and MZR. We also search for the projection of least scatter, compare KBSS galaxies to local galaxies, and quantify the anticorrelation between residual metallicity and SFR. Together, these tests comprehensively describe the $M_\star$-$Z_g$-SFR relationship in KBSS galaxies at $z\sim2.3$ in relation to the locally defined FMR.
\subsection{The Star-Forming Main Sequence}\label{sec:sfms}
\begin{figure}
    \centering
    \includegraphics[width=0.95\linewidth]{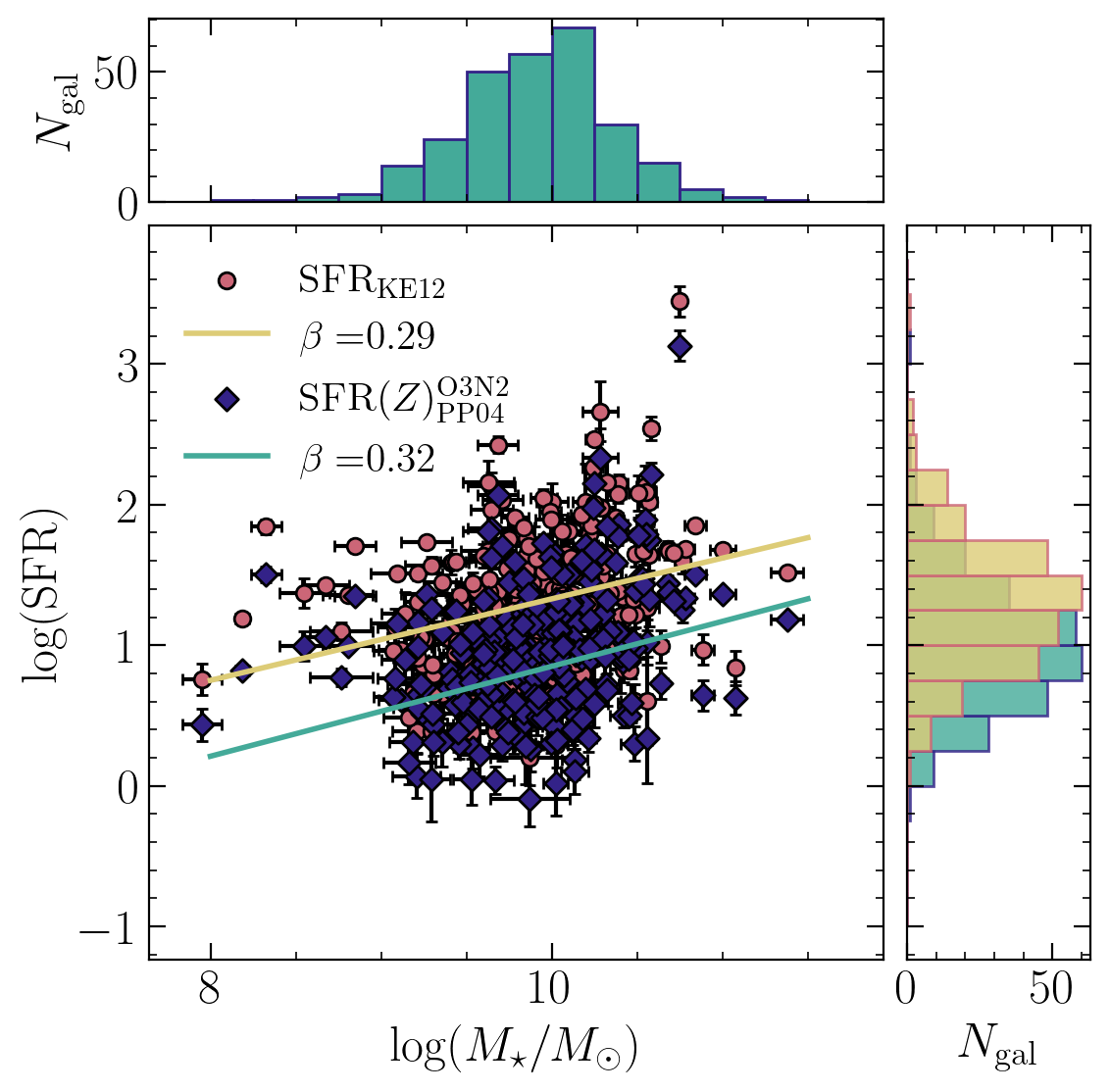}
    \caption{SFR vs. stellar mass for the KBSS O3N2 sample. $\mathrm{SFR}(Z)$ is shown by purple diamonds, with the regression line in teal and $\mathrm{SFR_{KE12}}$ is shown by pink circles, with the regression line in yellow. Histograms show the distributions of $\log(M_\star)$, $\log(\mathrm{SFR)_{KE12}}$ (yellow), and $\log(\mathrm{SFR}(Z))_{\mathrm{PP04}}^\mathrm{O3N2}$ (teal). Adopting $\mathrm{SFR}(Z)$ results in a SFMS with a lower normalization ($\mathrm{SFR}_{10}$) and a slightly steeper slope ($\beta$). }
    \label{fig:sfms_hists}
\end{figure}
\par A consequence of adopting a metallicity-dependent SFR conversion factor is a change in the star forming main sequence (SFMS). For each SFR method, we fit a power law to the data with the form 
\begin{equation}\label{sfms_eq}
    \log(\mathrm{SFR}) = \beta \times (\log(M_\star)-10) + \log(\mathrm{SFR}_{10})
\end{equation}
where $\beta$ is the slope of the SFMS and $\mathrm{SFR}_{10}$ is the median SFR for a galaxy with a stellar mass of $10^{10} M_\odot$. Here and throughout the paper, we use the \textsc{Python} package \textsc{Linmix} to simultaneously estimate intrinsic scatter and fit a regression line. This package is adapted from the IDL routine described in \cite{2007Kelly} and employs a Bayesian method designed to handle data with uncertainties in both the dependent and independent variables and an intrinsic level of scatter.
\par As can be seen in Figure \ref{fig:sfms_hists}, adopting SFR$(Z)$ results in a SFMS that is offset towards lower SFR at fixed $M_\star$ \emph{and} has a slightly steeper slope. The scale of the change is dependent on the strong-line calibration being used to derive oxygen abundance, and therefore, the iron abundance. Calibrations that result in the steepest MZR slopes (such as \citetalias{2018Bian}) are associated with a marginally steeper SFMS with a lower normalization. 
\par As is the case for all emission-line galaxy surveys, KBSS preferentially observes highly star-forming galaxies. Additionally, we impose SNR limits on the detection of H$\alpha$, therefore, the slope differs from the shapes reported in the literature using ``mass-selected" samples \citep[e.g.,][]{2014Whitaker, 2022Leja}. Depending on the choice of strong-line calibration, we recover $0.29\leq\beta\leq0.34$, while samples at similar redshifts have been found to have SFMS slopes of 0.91 ($\log(M_\star)<10.2$) and 0.67 \citep[$\log(M_\star)>10.2$][]{2014Whitaker}. The slopes we recover for KBSS galaxies are robust to outlier detection methods: when we use both random sample consensus (RANSAC) and jack-knifing, the slope remains statistically similar. When adopting $\mathrm{SFR_{KE12}}$, the normalization of our SFMS ($\log(\mathrm{SFR_{10}})=1.33$) is comparable to the SFR at $\log(M_\star)=10$ predicted by \citet[][$\log(\mathrm{SFR}_{10})=1.44$]{2014Whitaker}. However, adopting $\mathrm{SFR}(Z)$ leads to a significantly lower normalization ($\log(\mathrm{SFR}_{10})\sim0.85$). The SFR$(Z)$ SFMS is well-aligned with the SFMS predicted at $z\sim2$ by hydrodynamical simulations \citep{2014Torrey, 2019Donnari}, which tend to be offset towards a slightly lower SFR at fixed stellar mass when compared to the observed SFMS. Considering a metallicity dependence may therefore result in significantly lower normalizations of the SFMS in mass-selected samples.

\subsection{The Mass-Metallicity Relation}\label{sec:mzr}
\begin{figure*}
    \centering
    \includegraphics[width=1.0\linewidth]{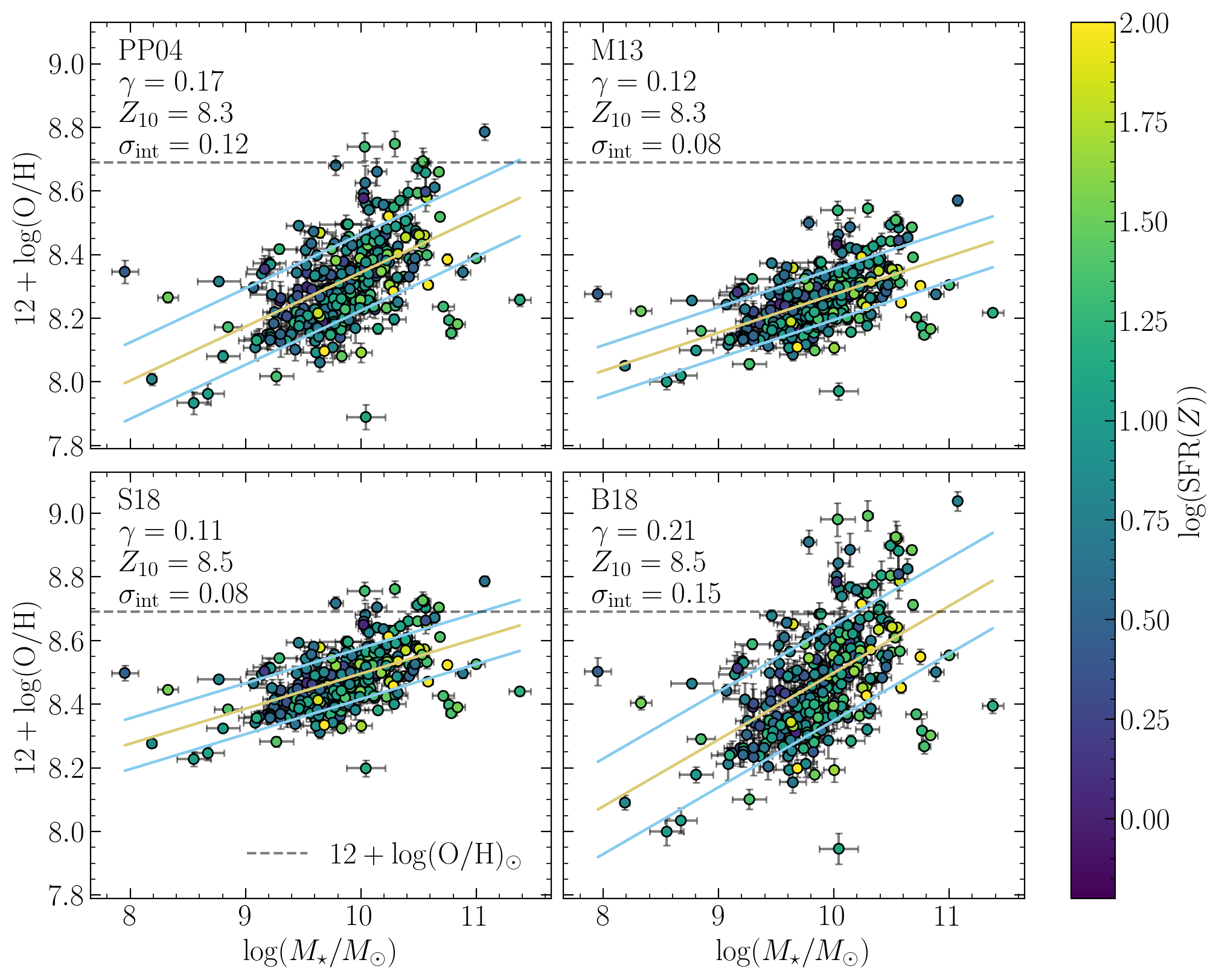}
    \caption{Gas-phase oxygen abundance vs. stellar mass for the KBSS O3N2 sample. The regression line is shown in yellow, intrinsic scatter is shown in light blue, individual galaxies are color-coded by $\log(\mathrm{SFR}(Z))$, and $12+\log(\mathrm{O/H})_\odot=8.69$ is shown by a gray dashed line. Each panel uses a different O3N2 calibration. Clockwise from top left: \citetalias{2004Pettini}, \citetalias{2013Marino}, \citetalias{2018Bian}, and \citetalias{2018Strom}. The slope ($\gamma$), normalization ($Z_{10}$), and scatter ($\sigma_\mathrm{int}$) of the MZR differ significantly depending on the chosen strong-line calibration.}
    \label{fig:o3n2_mzrs}
\end{figure*}

\begin{table}[]
    \centering
        \caption{Summary of MZR properties for different strong-line calibrations. We report the slope ($\gamma$), normalization ($Z_{10}$), intrinsic scatter ($\sigma_{\mathrm{int}}$), and the associated uncertainties. }
    \begin{tabular}{c|c c c}
    \hline
    \hline
         Calibration &  $\gamma$ & $Z_{10}$ & $\sigma_{\mathrm{int}}$ \\ \hline\hline
         \citetalias{2004Pettini}, O3N2 & $0.17 \pm 0.02$ & $8.344   \pm 0.008$ & $0.12 \pm 0.04$ \\
         \citetalias{2013Marino}, O3N2 & $0.12 \pm 0.01$ & $8.275 \pm 0.005$ & $0.08 \pm 0.03$\\
         \citetalias{2018Bian}, O3N2 & $0.21 \pm 0.02$ & $8.499 \pm 0.009$ & $0.15 \pm 0.05$\\
         \citetalias{2018Strom}, O3N2 & $0.11 \pm 0.01$ & $8.496 \pm 0.005$ & $0.08 \pm 0.02$ \\ \hline
         \citetalias{2013Marino}, N2 & $0.14 \pm 0.01$ & $8.378 \pm 0.006$ & $0.10 \pm 0.03$\\
         \citetalias{2014Steidel}, N2 & $0.11 \pm 0.01$ & $8.338 \pm 0.005$ & $0.08 \pm 0.02$\\
         \citetalias{2018Bian}, N2 & $0.14 \pm 0.02$ & $8.436 \pm 0.007$ & $0.11 \pm 0.03$ \\
         \citetalias{2018Strom}, N2 & $0.10 \pm 0.01$ & $8.504 \pm 0.005$ & $0.07 \pm 0.02$ \\\hline\hline
    \end{tabular}

    \label{tab:mzr_summary}
\end{table}

\par The eight different strong-line calibrations produce significantly different MZR slopes, scatters, and normalizations (see Figure \ref{fig:o3n2_mzrs}). For each calibration, we fit a power-law in the form 
\begin{equation}\label{mzr_powerlaw}
    12+\log(\mathrm{O/H})=\gamma\times(\log(M_\star)-10)+Z_{10} +\mathcal{N}(0, \sigma_\mathrm{int})
\end{equation}
In the case of O3N2-based oxygen abundance, the slope can be as high as $\gamma=0.21\pm0.02$ when using the \citetalias{2018Bian} calibration or as low as $\gamma=0.11\pm0.01$ when using either the \citetalias{2013Marino} or \citetalias{2018Strom} calibration. The normalization varies by 0.224 dex between different calibrations, where the photoionization model-based \citetalias{2018Bian} calibration produces the highest normalization ($Z_{10}=8.499\pm0.009$) and the \citetalias{2013Marino} calibration produces the lowest normalization ($Z_{10}=8.275\pm0.005$). N2-based MZRs tend to have slightly shallower slopes and less scatter than O3N2-based MZRs. Across both strong-line ratios, calibrations with a larger $|B|$ value (see Table \ref{tab:calibrations}) result in steeper MZRs, with more scatter. For a summary of our MZR results, see Table \ref{tab:mzr_summary} and Figure \ref{fig:o3n2_mzrs}. 
\par Testing calibrations that result in different underlying MZR shapes allows us to determine whether the detection of the FMR is sensitive to the MZR slope and scatter, as well as the choice of line ratio. Other studies have chosen to look for the FMR in line ratio space \citep[e.g.,][]{2018Sanders}, which has a similar effect of ensuring the detection of the FMR is not reliant on the MZR shape. 

\begin{figure*}
    \centering
    \includegraphics[width=1.0\linewidth]{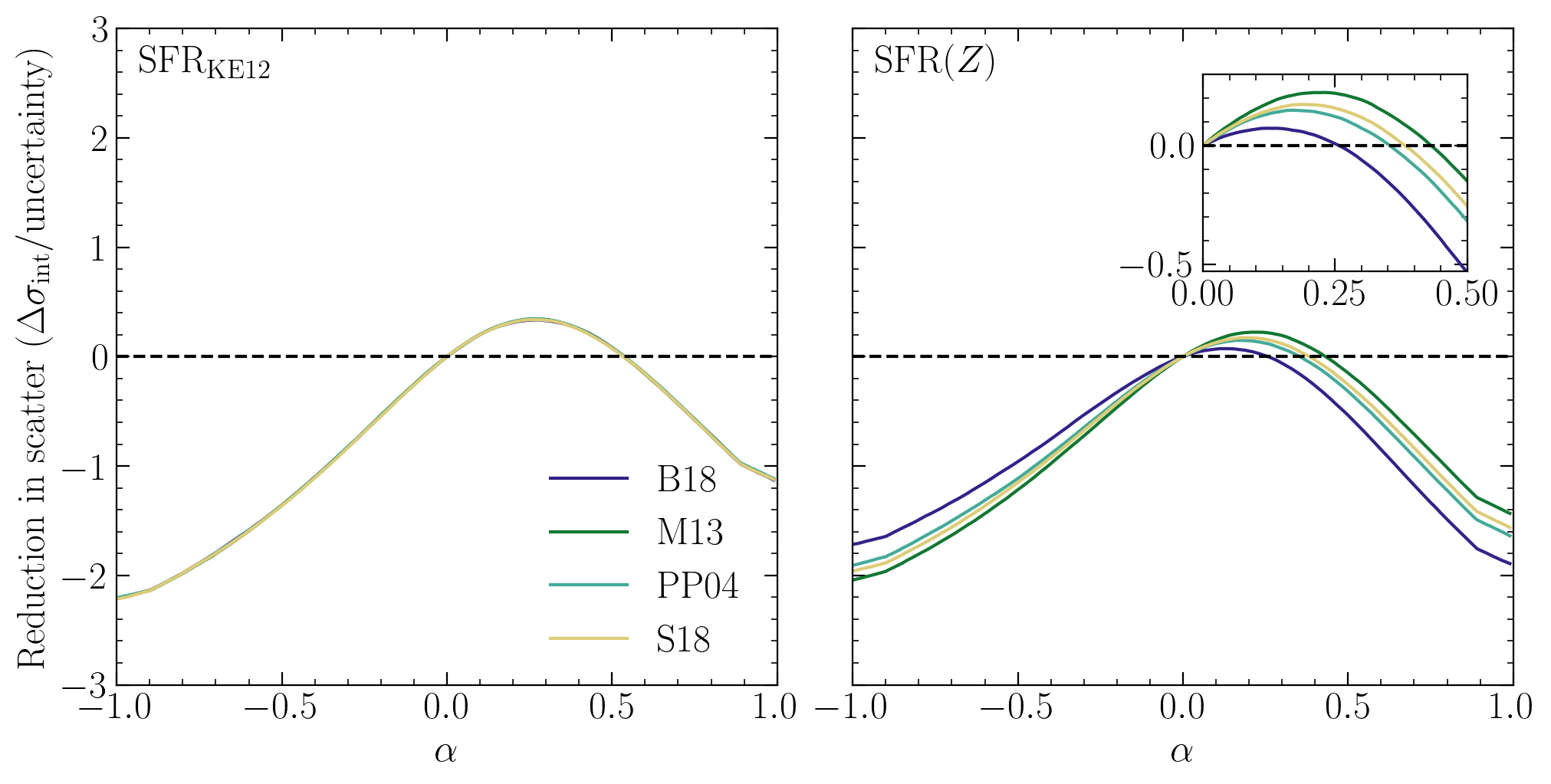}
    \caption{Reduction in intrinsic scatter vs. $\alpha$ for the four O3N2 strong-line calibrations and using $\mathrm{SFR_{KE12}}$ (left panel) and $\mathrm{SFR}(Z)$ (right panel). The ordinate is specifically defined as the reduction in intrinsic scatter ($\Delta\sigma_\mathrm{int}$), relative to the MZR scatter, divided by the uncertainty on the intrinsic scatter. The inset panel zooms in to $0\leq\alpha\leq0.5$ to show the effect of adopting SFR$(Z)$. S\deleted{ince none of the solid curves (each corresponding to a different calibration) peak above 1, s}catter is not significantly reduced at any value of $\alpha$. See Appendix \ref{app:n2} for results using N2 calibrations.
    } 
    \label{fig:o3n2_alpha}
\end{figure*}



\subsection{Parametric Method}\label{sec:alpha_method}
\par \cite{2010Mannucci} originally described the FMR as a three-dimensional curved surface in gas phase oxygen abundance, stellar mass, SFR space. Different two-dimenstional projections of this surface can be represented by the parameter $\mu_\alpha\equiv\log(M_\star)-\alpha\log(\mathrm{SFR})$ where $\alpha$ has a value between $-1$ and $1$, with $\alpha=0$ corresponding to the MZR, $\alpha=1$ corresponding to a direct relationship between metallicity and sSFR, and intermediate values corresponding to intermediate projections. \cite{2010Mannucci} found the two-dimensional projection of the three-dimensional FMR that minimizes scatter at $\alpha_{\min}=0.32$. This parametric approach has been applied with varying results at a range of redshifts \citep{2012Yates, 2013Andrews, 2020Curti, 2021Sanders} and it remains an open question whether the projection of least scatter is redshift-invariant \citep{2024Garcia_a}. 
\par We follow a similar method to \cite{2010Mannucci} and define the abscissa by the variable $\mu_\alpha\equiv\log(M_\star)-\alpha\log(\mathrm{SFR})$. We sample values of $\alpha$ between $-1$ and 1 in steps of 0.01. At each step, we fit a a regression line to the data in the form 
\begin{equation}
    12+\log(\mathrm{O/H}) = \gamma\times\mu_{\alpha} + Z_0+\mathcal{N}(0, \sigma_\mathrm{int})
\end{equation}
and estimate the instrinsic scatter using \textsc{Linmix}.
Since \textsc{Linmix} results have some stochasticity, we apply a boxcar smoothing function to the intrinsic scatter values, smoothing over a window of $\pm0.1$ in $\alpha$. 
\par Across all strong-line calibrations, SFR calculation methods, and means of estimating intrinsic scatter, we find no statistically significant reduction in scatter at $\alpha\neq 0$. Figure \ref{fig:o3n2_alpha} shows the results of the parametric analysis using the O3N2 sample and either $\mathrm{SFR_{KE12}}$ (left panel) or $\mathrm{SFR}(Z)$ (right panel). The ordinate is defined as the reduction in scatter, measured relative to the scatter of the MZR ($\Delta\sigma_\mathrm{int}$) and divided by the uncertainty on intrinsic scatter. In the left panel of Figure \ref{fig:o3n2_alpha}, the solid curves (where each color represents a different strong-line calibration) peak around $\alpha\sim0.3$ and in the right panel, the curves peak between $0.1\lesssim\alpha\lesssim0.2$, showing that scatter is ostensibly minimised at $\alpha\neq0$. However, the uncertainty on the intrinsic scatter is such that any reduction in scatter is not statistically significant It is worth noting that while all of the curves in the left panel of Figure \ref{fig:o3n2_alpha} overlap, the absolute reduction in scatter does differ by calibration. Calibrations that more heavily weight the line ratio (larger $B$ in Table \ref{tab:calibrations}) show a larger reduction in scatter, albeit with a larger uncertainty on the scatter. Hence, when we normalise the reduction in scatter by the uncertainty, all of the curves collapse onto one another. 
\par We find a similar result using N2-based calibrations, although the relative reduction in scatter is smaller and peaks at a lower value of $\alpha$. Given that O3N2 has been shown to produce oxygen abundances in better agreement with direct $T_e$ abundances in KBSS \citep{2014Steidel} and the consistency between our O3N2- and N2-based results, we do not include further results using the N2 calibrations in the main body of the paper. For further detail regarding the N2 calibration results, we refer the reader to Appendix \ref{app:n2}. 
\par As a point of comparison, we also employ the same method used in the original \cite{2010Mannucci} paper, where we consider the dispersion in $\log(\mathrm{O/H})$ about the best-fit MZR. To estimate the uncertainty in the dispersion, we perturb each point in both $12+\log(\mathrm{O/H})$ and $\mu_{0.32}$ (the projection at which dispersion is minimized) 100 different times, drawing the perturbation from a normal distribution with a standard deviation equal to the measurement uncertainty. Then, we take the standard deviation of the simulated dispersion measurements as the uncertainty on the dispersion. Our results are unchanged when we measure total dispersion (uncorrected for observational uncertainty), as in \cite{2010Mannucci}, as opposed to intrinsic scatter. Therefore, we conclude that the discrepancy between our result and the \cite{2010Mannucci} result is not due to choice of method, but reflects intrinsic differences between local galaxy samples and KBSS galaxies. 

\subsection{Comparison to the Local FMR}\label{sec:am13_comp}
\par
\begin{figure*}
    \centering
    \includegraphics[width=1.0\linewidth]{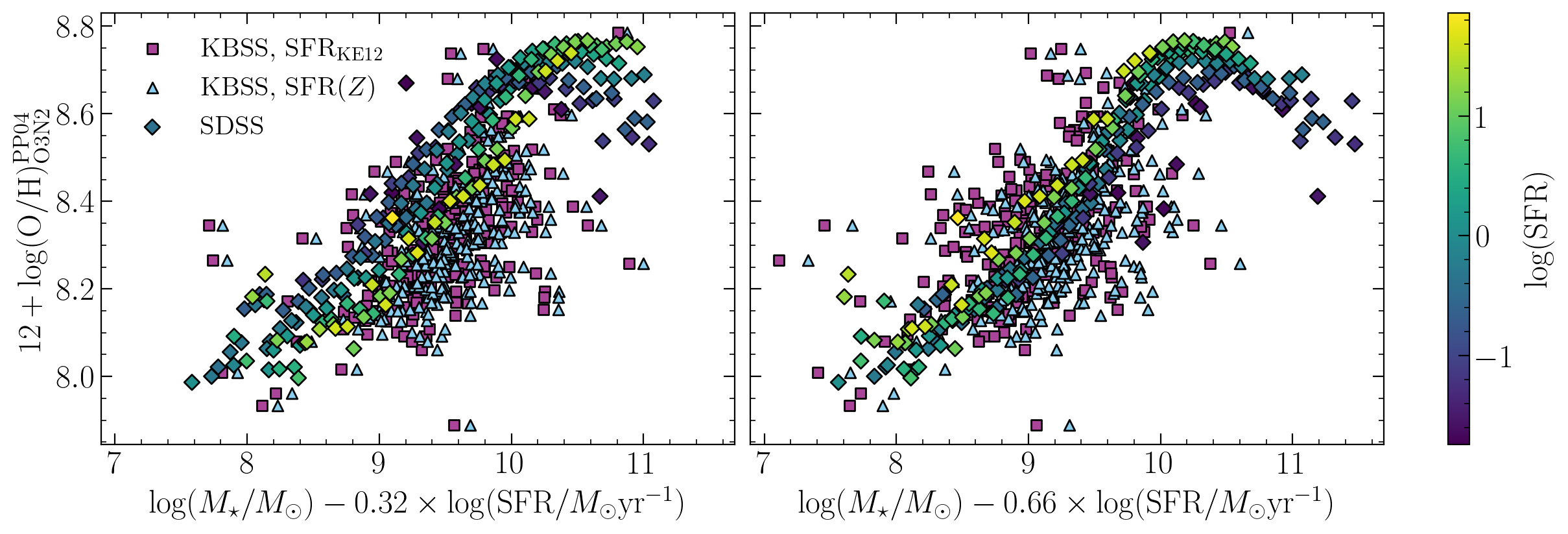}
    \caption{Oxygen abundance vs. $\mu_\alpha$ for SDSS galaxies \citep[diamonds, color-coded by SFR, from][]{2013Andrews} and KBSS galaxies (purple squares for $\mathrm{SFR_{KE12}}$ and light blue triangles for SFR$(Z)$). The left panel plots $\mu_{0.32}$ and the right panel plots $\mu_{0.66}$. On the left, KBSS galaxies are offset towards lower oxygen abundances at fixed $\mu_{0.32}$. On the right, KBSS galaxies and SDSS galaxies overlap, although scatter is not reduced for KBSS galaxies at this projection.}
    \label{fig:am13fmr}
\end{figure*}

 It is possible that KBSS galaxies lie on the FMR but in an area of $M_\star$-$Z_g$-SFR space where the dependence on SFR is weak and a significant reduction in scatter is too subtle to be measured. To determine if this is the case, we compare KBSS galaxies to $z\sim0$ galaxies that have been used to define the local FMR. Specifically, we use the stacked SDSS spectra from \cite{2013Andrews}, with $M_\star$ and SFR rescaled to a \cite{2003Chabrier} IMF. To allow for the most direct comparison, we consider $12+\log(\mathrm{O/H})$ calculated from O3N2 \citetalias{2004Pettini} for both KBSS and SDSS galaxies. For this calibration, \cite{2013Andrews} report that scatter is minimized at $\alpha_{\min}=0.32$, and we show this projection in the left panel of Figure \ref{fig:am13fmr}. However, there is still a noticeable SFR gradient in the \cite{2013Andrews} stacks at this projection, suggesting that it is perhaps not the optimal projection. In the right panel of Figure \ref{fig:am13fmr}, we use $\mu_{0.66}=\log(M_\star)-0.66\times\log(\mathrm{SFR})$ as the abscissa instead. \cite{2013Andrews} report that scatter is minimized at $\alpha_{\min}=0.66$ when using direct method $T_e$ abundances. As can be seen in Figure \ref{fig:am13fmr}, the trend in SFR is reduced at this projection. However, stacks with high SFR are pushed to low $\mu_{0.66}$ and have very large scatter from the otherwise tight series in oxygen abundance and $\mu_{0.66}$. These highly scattered points may explain why scatter is technically minimized at a lower value of $\alpha$, despite there still being a visible trend in SFR.

\par In the left panel of Figure \ref{fig:am13fmr}, KBSS galaxies are offset towards lower $12+\log(\mathrm{O/H})$ at fixed $\mu_{0.32}$ compared to the $z\sim0$ SDSS galaxies. The offset is slightly larger when using SFR$(Z)$ due to the bulk offset between $\mathrm{SFR_{KE12}}$ and SFR$(Z)$. In the right panel of Figure \ref{fig:am13fmr}, KBSS and SDSS appear to overlap.
\par However, we caution that an overlap in $\log(\mathrm{O/H})$-$\mu_{0.66}$ space does not necessarily imply that KBSS and SDSS fall on a shared fundamental plane in $M_\star$-$Z_g$-SFR space. In Section \ref{sec:othersamples}, we further discuss how our results differ from predictions associated with a redshift-invariant FMR plane, demonstrating that the overlap between KBSS and SDSS is not due to a shared plane.

\subsection{Non-Parametric Method}\label{sec:nonparam}
\par A complementary, non-parametric approach was proposed by \cite{2014Salim}, who did not adopt the \cite{2010Mannucci} parametrization of the FMR. Instead, \cite{2014Salim} quantified the anticorrelation between oxygen abundance and the offset of a galaxy from the star forming main sequence. \cite{2014Salim} argued that this method is more physically motivated, as it considers the relationship between $Z_g$ and the galaxy's residual specific star formation rate (sSFR), relative to the typical sSFR for a galaxy of the same mass, rather than absolute SFR, and does not require the surface in $M_\star$-$Z_g$-SFR to be a plane. Various non-parametric approaches have been employed to investigate the $M_\star$-$Z_g$-$\mathrm{SFR}$ relation, generally comparing residual oxygen abundance ($\Delta\log(\mathrm{O/H})$), defined as the difference between a galaxy's oxygen abundance and the expected oxygen abundance for a galaxy of the same stellar mass, with a third, SFR-based parameter \citep[e.g.,][]{2021Sanders, 2021Topping}.  The FMR does predict that $\Delta\log(\mathrm{O/H})$ will be anticorrelated with SFR, although it should be noted that an anticorrelation between $\Delta\log(\mathrm{O/H})$ and SFR is not necessarily consistent with the FMR plane defined by \cite{2010Mannucci}, an issue we discuss further in Section \ref{sec:injectionrecovery}. 

\par Here we consider two versions of the non-parametric approach: first, we quantify the correlation between residual oxygen abundance and absolute SFR. Residual oxygen abundance is calculated relative to the power law (Equation \ref{mzr_powerlaw}) parametrization of the MZR for each calibration (summarized in Table \ref{tab:mzr_summary}) such that $\Delta\mathrm{MZR}\equiv(12+\log(\mathrm{O/H})) - (\gamma\times(\log(M_\star)-10) + Z_{10})$. We then use \textsc{Linmix} to perform a linear regression on the residual oxygen abundance and SFR and report the Spearman's rank correlation coefficient, $\rho$, and $p$-value. 

\par
\begin{figure}
    \centering
    \includegraphics[scale=0.6]{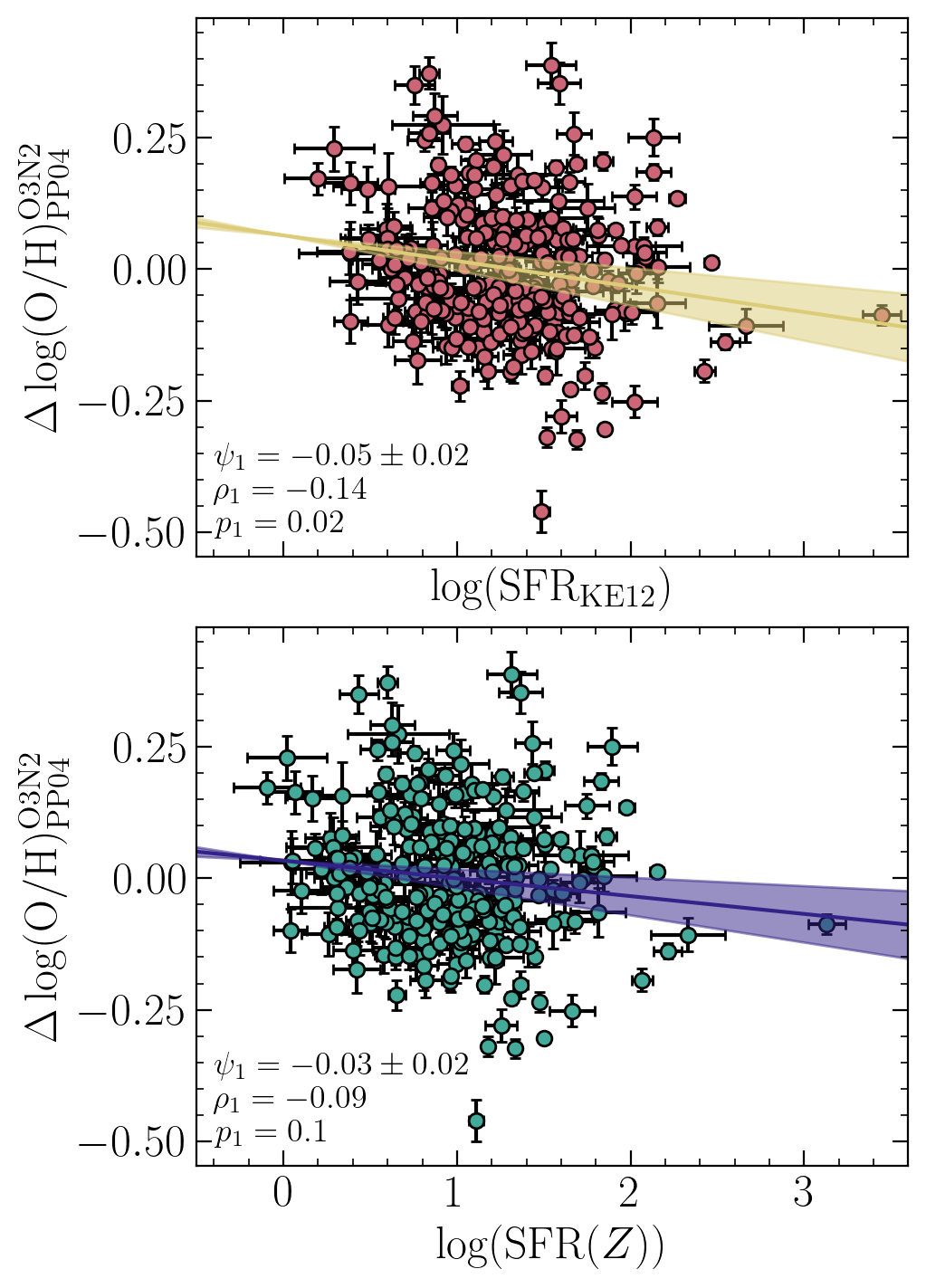}
    \caption{\emph{Top}: $\Delta\log(\mathrm{O/H})$ vs. $\log(\mathrm{SFR_{KE12}})$ for the KBSS O3N2 sample. \emph{Bottom}: as above, but using $\log(\mathrm{SFR}(Z))$. There is a weak and marginally significant anticorrelation between $\Delta\log(\mathrm{O/H})$ and SFR in the top panel, and no significant correlation in the bottom panel. The trend is seen across calibrations, here we show the results for the \citetalias{2004Pettini} calibration.}
    \label{fig:mzr_res}
\end{figure}

\begin{figure}
    \centering
    \includegraphics[scale=0.6]{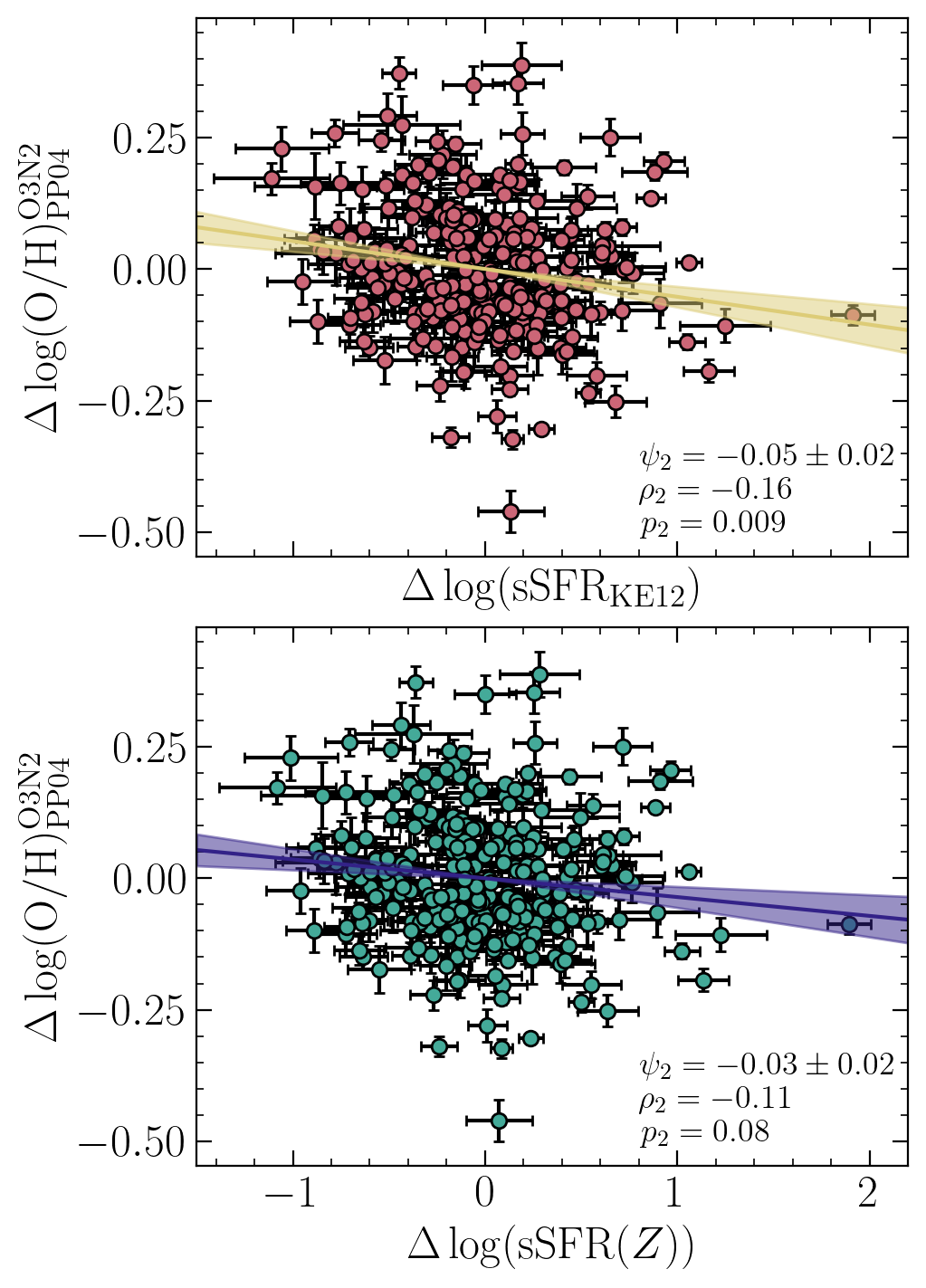}
    \caption{\emph{Top}: $\Delta\log(\mathrm{O/H})$ vs. $\Delta\mathrm{sSFR_{KE12}}$ for the KBSS O3N2 sample. \emph{Bottom}: as above, but using $\Delta\mathrm{sSFR}(Z)$.
    A weak and marginally significant $\Delta\log(\mathrm{O/H})$-$\Delta$sSFR anticorrelation is observed only when using $\mathrm{SFR_{KE12}}$. As in Figure \ref{fig:mzr_res}, we use the \citetalias{2004Pettini} calibration, but this result is observed across calibrations.}
    \label{fig:res_res}
\end{figure}

 Secondly we quantify the correlation between $\Delta\log(\mathrm{O/H})$ and the relative sSFR, following the method outlined by \cite{2014Salim}. First, we characterise the mass-sSFR relationship for KBSS galaxies, and use the regression line to calculate the residual sSFR for each galaxy ($\Delta\mathrm{sSFR}\equiv \mathrm{sSFR}-\langle\mathrm{sSFR}(M_\star)\rangle$). Similarly, we use \textsc{Linmix} to perform a linear regression and report $\rho$ and $p$. The resulting $\Delta\log(\mathrm{O/H})$-SFR and $\Delta\log(\mathrm{O/H})$-$\Delta$sSFR correlations can be seen in Figures \ref{fig:mzr_res} and \ref{fig:res_res} respectively.


\begin{table*}[]
    \centering
    \caption{The slope ($\psi_1$, $\psi_2$), Spearman's rank correlation coefficient ($\rho_1$, $\rho_2$), $p$-value ($p_1$, $p_2$) for the $\Delta\mathrm{MZR}_{\mathrm{O3N2}}$-SFR and $\Delta\mathrm{MZR}_{\mathrm{O3N2}}$-$\Delta$sSFR anticorrelations respectively.}
    \begin{tabular}{cc|ccc|ccc}
        \hline\hline
        \multirow{2}{*}{Calibration} & \multirow{2}{*}{SFR} & \multicolumn{3}{c|}{$\Delta\log(\mathrm{O/H})$-SFR} & \multicolumn{3}{c}{$\Delta\log(\mathrm{O/H})$-$\Delta$sSFR} \\ 
         &                      & $\psi_1$    & $\rho_1$    & $p_1$   & $\psi_2$       & $\rho_2$       & $p_2$      \\\hline\hline
         \citetalias{2004Pettini} & \citetalias{2012Kennicutt} & $-0.05 \pm 0.02$ & $-0.14$ & $0.02$ & $-0.05\pm0.02$ & $-0.16$& 0.009 \\
         \citetalias{2004Pettini} & Z & $-0.03 \pm 0.02$ & $-0.09$ & $0.1$ & $-0.03\pm0.02$& $-0.11$& 0.08\\
         \citetalias{2013Marino} & \citetalias{2012Kennicutt} & $-0.04 \pm 0.01$ & $-0.15$ & $0.01$ & $-0.04\pm0.01$ & $-0.16$ & 0.009\\
        \citetalias{2013Marino} & Z & $-0.03 \pm 0.01$ & $-0.12$ & $0.05$ & $-0.03\pm0.01$ & $-0.13$ & 0.03\\
        \citetalias{2018Bian} & \citetalias{2012Kennicutt} & $-0.06 \pm 0.02$ & $-0.14$ & $0.02$ & $-0.06\pm0.02$ & $-0.16$ & 0.009\\
        \citetalias{2018Bian} & Z & $-0.03 \pm 0.02$ & $-0.06$ & $0.32$ & $-0.03\pm0.02$ & $-0.08$ & 0.2\\
        \citetalias{2018Strom} & \citetalias{2012Kennicutt} & $-0.03 \pm 0.01$ & $-0.14$ & $0.02$ & $-0.03\pm0.01$ & $-0.16$ & 0.009\\
        \citetalias{2018Strom} & Z & $-0.02 \pm 0.01$ & $-0.09$ & $0.12$ & $-0.02\pm0.01$ & $-0.11$ & 0.07\\
        \hline\hline
    \end{tabular}
    \label{tab:dmzr_dssfr}
\end{table*}

\par Across calibrations, we find a weak but significant ($p<0.02$) anticorrelation between residual oxygen abundance ($\Delta\log(\mathrm{O/H})$) and $\mathrm{SFR_{KE12}}$. We also observe a weak but significant ($p<0.009$) anticorrelation between $\Delta\log(\mathrm{O/H})$ and $\Delta\mathrm{sSFR_{KE12}}$. See Table \ref{tab:dmzr_dssfr} for a full list $p$-values and Spearman $\rho$ coefficients by calibration and SFR calculation. While the $\Delta\log(\mathrm{O/H})$-$\Delta$sSFR slopes recovered for KBSS (yellow and purple lines in Figure \ref{fig:res_res}) are significantly shallower than what has been found at $z\sim2.3$ by \citet[][$\psi_2=-0.14$, see also Appendix \ref{app:mosdef}]{2018Sanders}, the scatter in KBSS galaxies is large enough that \cite{2018Sanders} slope could be consistent with KBSS. 
\par However, when adopting $\mathrm{SFR}(Z)$, the correlation is weaker and only marginally significant or consistent with a null result. The difference between the the results for the two SFR methods is easily understood. While the principal effect of adopting $\mathrm{SFR}(Z)$ is a systematic shift towards lower SFR for all galaxies, the effect is larger for more metal-poor galaxies and smaller for more metal-rich galaxies (see Figure \ref{fig:sfr_hist}). Therefore, a set of galaxies which show an anticorrelation between residual oxygen abundance and $\mathrm{SFR_{KE12}}$ will have this correlation weakened by the adoption of $\mathrm{SFR}(Z)$. Conversely, a more metal-poor galaxy will produce more $\mathcal{L}_{\mathrm{H}\alpha}$ than a more metal-rich galaxy with the same SFR and stellar mass. Assuming a constant conversion factor would, however, suggest that the more metal-poor galaxy is more highly star-forming, introducing an artificial anticorrelation between oxygen abundance and SFR. By the same logic, we can understand why the calibration with the steepest MZR slope, \citetalias{2018Bian}, results in the largest suppression of the anticorrelation when switching from $\mathrm{SFR_{KE12}}$ to $\mathrm{SFR}(Z)$. 
\par The $p$-values for both the $\Delta\log(\mathrm{O/H})$-SFR$(Z)$ and $\Delta\log(\mathrm{O/H})$-$\Delta$sSFR$(Z)$ anticorrelations are such that we cannot rule out a $M_\star$-$Z_g$-SFR$(Z)$ relation. Moreover, the strength of the correlation is clearly dependent on the MZR slope, the ``true" value of which remains poorly constrained. Thus, it remains uncertain whether a $M_\star$-$Z_g$-SFR$(Z)$ relation exists in KBSS galaxies, but it is evident that a constant SFR conversion factor (such as the \citetalias{2012Kennicutt} conversion, although this trend will be seen for any constant conversion factors) can result in an overestimate of the correlation strength. 

\par N2-based calibrations yield similar results: $\Delta\log(\mathrm{O/H})$ is marginally and weakly anticorrelated with both SFR and $\Delta$sSFR. Further detail and figures illustrating this can be found in Appendix \ref{app:n2}. For both $\mathrm{SFR_{KE12}}$ and SFR$(Z)$, the anticorrelation with $\Delta\log(\mathrm{O/H})_\mathrm{N2}$ is weaker than the anticorrelation with $\Delta\log(\mathrm{O/H})_\mathrm{O3N2}$ (cf. Tables \ref{tab:dmzr_dssfr} and \ref{tab:n2_res}). The discrepancy highlights that strong-line calibrations are not self-consistent and are sensitive to metallicity in different ways. At fixed stellar mass, N2 is less strongly correlated with SFR than O3N2. Any correlation between a strong line ratio and SFR may be indicative of a correlation between metallicity and SFR, or an independent correlation between the line ratio and SFR. Our analysis cannot differentiate between these two situations, only show that the strength of the correlation varies between line ratios. 

\section{Discussion}\label{sec:discussion}

\subsection{Reconciling Results From Parametric and Non-Parametric Methods}\label{sec:injectionrecovery}
\par At face value, it would seem contradictory  to observe a $M_\star$-$Z_g$-$\mathrm{SFR}$ relation without a corresponding projection of least scatter. Given that the presence of the FMR is sometimes reported based on $\Delta\log(\mathrm{O/H})$-SFR or $\Delta\mathrm{MZR}$-$\Delta\mathrm{sSFR}$ correlations \citep[e.g.,][]{2021Topping}, it's important to understand how a $M_\star$-$Z_g$-$\mathrm{SFR}$ relation may or may not imply a reduction in intrinsic scatter. 

\par To explore the relationship between the parametric and non-parametric methods, we use an injection-recovery test, where we input simulated data based on an artificial $M_\star$-$Z_g$-SFR relationship into the parametric analysis method. For each realization of the data, we measure the reduction in scatter for $0\leq\alpha\leq1$.

\par The injected data has the same mass and SFR distribution as the KBSS O3N2 sample, using $\mathrm{SFR_{KE12}}$. Each data point is assigned a residual metallicity based on a constructed $\Delta\log(\mathrm{O/H})$-SFR relationship, which takes the form of
\begin{equation}\label{res_eq}
    \Delta\log(\mathrm{O/H}) = \psi_1 \times \log(\mathrm{SFR}) + \mathcal{N}(0, \sigma_\mathrm{int}) + \mathcal{N}(0, \sigma_\mathrm{meas})
\end{equation}
where $\psi_1$ is the slope, $\sigma_\mathrm{int}$ is intrinsic scatter in the $\Delta\log(\mathrm{O/H})$-SFR relationship, and $\sigma_\mathrm{meas}$ is the measurement uncertainty. We assume a constant value of $\sigma_\mathrm{meas}=0.03$, equal to the median measurement uncertainty on $\Delta\log(\mathrm{O/H})$ in the O3N2 sample using the \citetalias{2004Pettini} calibration. Without loss of generality, we define $\Delta\log(\mathrm{O/H})=0$ at $\log(\mathrm{SFR})=0$. The value of $\Delta\log\mathrm{(O/H)}$ is then added to the value of $\log\mathrm{(O/H)}$ predicted by the power law form of the MZR (Equation \ref{mzr_powerlaw}), given the associated stellar mass. $\psi_1$ and $\sigma_\mathrm{int}$ are both varied to produce the simulated data sets. For each realization of the data, we then use the same parametric method described in Section \ref{sec:alpha_method} to find the value of $\alpha$ that minimizes scatter and measure the reduction in scatter, relative to the MZR ($\alpha=0$).  
\begin{figure}
    \centering
    \includegraphics[width=1.0\linewidth]{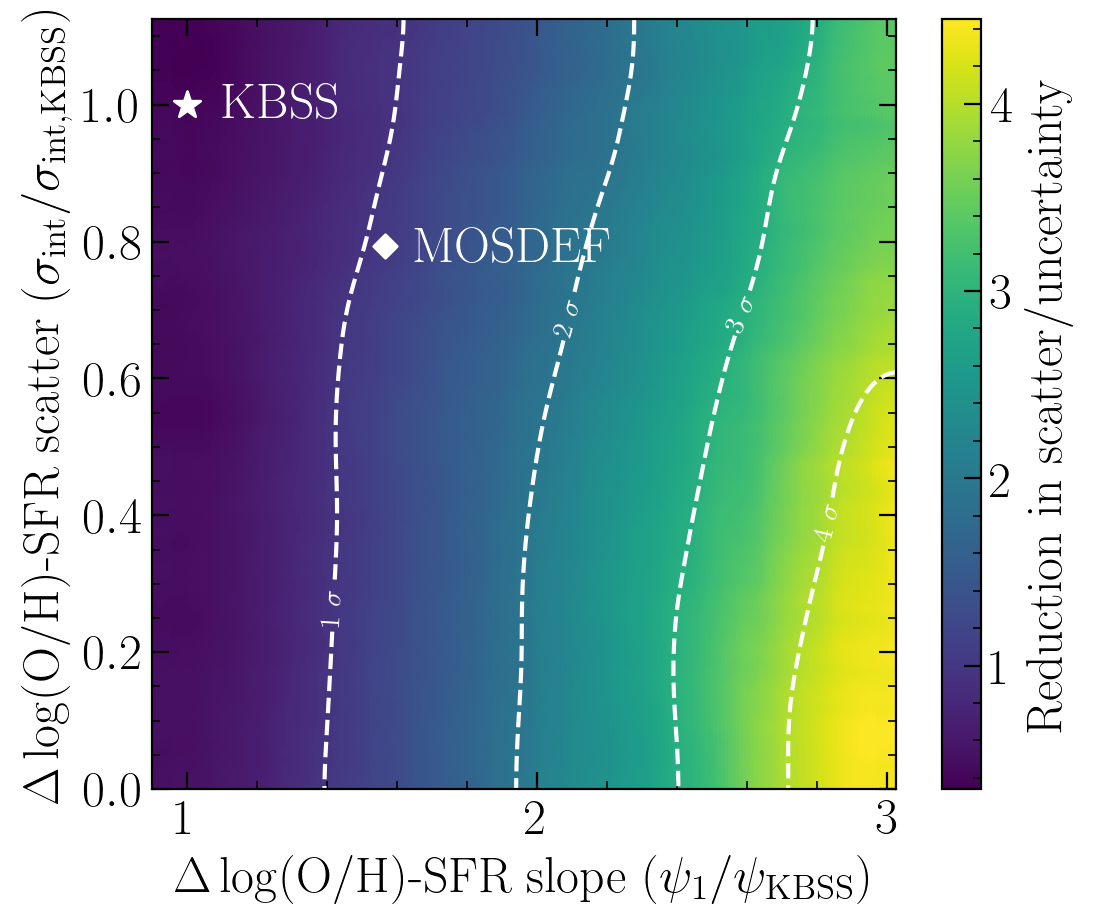}
    \caption{$\Delta\log(\mathrm{O/H})$-SFR scatter vs. $\Delta\log(\mathrm{O/H})$-SFR slope of the simulated datasets, color-coded by reduction in scatter at $\alpha_\mathrm{min}$. Slope and scatter have been normalized by $\psi_\mathrm{KBSS}=-0.05$ and $\sigma_\mathrm{int, KBSS}=0.13$ respectively (shown by the white star), and reduction in FMR scatter is normalized by the uncertainty ($\sigma$) on the FMR scatter. The MOSDEF slope and scatter are shown by white diamond. A significant reduction in scatter is only reliably measured at steeper $\psi_1$, showing that conclusions drawn from parametric and non-parametric methods are not necessarily consistent.}
    \label{fig:slope_scatter_grid}
\end{figure}
\par Figure \ref{fig:slope_scatter_grid} shows the results of our injection-recovery test. Each pixel represents a simulated dataset which was constructed according to Equation \ref{res_eq} with a different value of $\psi_1$ and $\sigma_\mathrm{int}$. The color of each pixel shows the reduction in intrinsic scatter, relative to the MZR scatter, divided by the uncertainty. A significant reduction ($>3\sigma$) in scatter can only be observed alongside a significantly steeper $\Delta\log(\mathrm{O/H})$-SFR relation, regardless of the intrinsic scatter in the relationship. In fact, $\psi_1$ must be at least 2.4 times steeper than $\psi_\mathrm{KBSS}$ before a significant reduction in scatter is measured. Hence, an anticorrelation between residual metallicity and SFR is is only associated with a significant reduction in scatter if the anticorrelation is much stronger than what we measure in KBSS. Indeed, the MOSDEF sample shows a stronger $\Delta\log(\mathrm{O/H})$-SFR anticorrelation (see Figure \ref{fig:eres_res_mosdef} and Appendix \ref{app:mosdef}) with less scatter. Values of $\psi_1$ and $\sigma_\mathrm{int}$ for MOSDEF are marked on Figure \ref{fig:slope_scatter_grid} by a white diamond. At these values, our injection recovery tests predict a marginal, $\sim1.2\sigma$ reduction in scatter. Applying our parametric analysis method to MOSDEF (see Figure \ref{fig:mosdef_alpha}) yields a $\sim1.5\sigma$ reduction in scatter at $\alpha_{\min}$. Similarly, the SDSS stacks from \cite{2013Andrews} show a significant reduction in scatter at $\alpha_{\min}$ and have much steeper $\Delta\log(\mathrm{O/H})$-SFR slope ($\psi_1=-0.22\pm{0.02}$, over four times steeper than KBSS).
\par The same conclusion can be reached analytically. Assuming there is a significant $\Delta\log(\mathrm{O/H})$-SFR anticorrelation, total MZR scatter ($\sigma_\mathrm{tot}$) can be decomposed into three components:
\begin{equation}
    \sigma_\mathrm{tot}^2 = (\psi_1\sigma_\mathrm{SFR})^2 + \sigma_\mathrm{int}^2 + \sigma_\mathrm{meas}^2
\end{equation}
where $\sigma_\mathrm{meas}$ is scatter due to measurement error in $\log(\mathrm{O/H})$, $\sigma_\mathrm{int}$ is the intrinsic scatter in the $\Delta\log(\mathrm{O/H})$-SFR correlation (ordinate of Figure \ref{fig:slope_scatter_grid}), $\psi_1$ is the slope of the $\Delta\log(\mathrm{O/H})$-SFR correlation (abscissa of Figure \ref{fig:slope_scatter_grid}), and $\sigma_\mathrm{SFR}$ is the standard deviation of $\log(\mathrm{SFR})$. When we estimate the total intrinsic scatter using \textsc{Linmix}, we are estimating the combination of $\psi_1\sigma_\mathrm{SFR}$ and $\sigma_\mathrm{int}$. In the case of very shallow $\Delta\log(\mathrm{O/H})$-SFR anticorrelations, as is seen in KBSS, the total intrinsic scatter is dominated by the scatter in $\Delta\log(\mathrm{O/H})$-SFR. The parametric approach solves for the value of $\alpha$ which minimizes the contribution of $\psi_1\sigma_\mathrm{SFR}$ to the total intrinsic scatter (i.e., when $\alpha_{\min}=\psi_1/\gamma$). However, in KBSS, $\psi_1\sigma_\mathrm{SFR}$ is very small to begin with, resulting in no significant reduction in scatter. Indeed in Figure \ref{fig:slope_scatter_grid}, we only begin to see a significant reduction in scatter when $\psi_1$ is much steeper than $\psi_\mathrm{KBSS}$.
\subsection{Comparison to Other Samples}\label{sec:othersamples}
\par Figure \ref{fig:am13fmr} shows that depending on the choice of $\mu_\alpha$, KBSS may or may not appear to overlap with local galaxies. If we choose $\mu_{0.32}$, which \cite{2013Andrews} states reduces scatter when using the \citetalias{2004Pettini} calibration, KBSS galaxies are offset from the locally-defined FMR towards lower oxygen abundances at fixed $\mu_{0.32}$. But, if we choose $\mu_{0.66}$, which reduces scatter of $z\sim0$ galaxies when using direct-method $T_e$ abundances \citep{2013Andrews}, KBSS galaxies appear to fully overlap with SDSS stacks.
\par Although KBSS and SDSS do overlap in the right panel of Figure \ref{fig:am13fmr}, this does not necessarily mean that they occupy the same plane in $M_\star$-$Z_g$-SFR space. Indeed, we can see from Figure \ref{fig:o3n2_alpha} that scatter is nominally increased for KBSS galaxies at $\alpha=0.66$. $\mu_{0.66}$ may not be the projection of least scatter for KBSS galaxies, just a projection at which KBSS and SDSS galaxies overlap. If KBSS and SDSS were on a fundamental, redshift-invariant plane in $M_\star$-$Z_g$-SFR space, then we would expect scatter to be minimized at the same value of $\alpha$ at all redshifts.
\par However, as has been noted in Sections \ref{sec:alpha_method} and \ref{sec:injectionrecovery}, the uncertainty on scatter is such that we don't see a significant reduction in scatter at \emph{any} value of $\alpha$. Therefore, it's difficult to argue that we have found a different value of $\alpha_{\min}$ for KBSS galaxies compared to $\alpha=0.66$. Additionally, the increase in KBSS galaxy scatter at $\alpha=0.66$ is also not statistically significant. We can attempt to analytically determine $\alpha_{\min}$: as was noted in Section \ref{sec:injectionrecovery}, $\alpha_{\min}$ minimizes the contribution of SFR to the total scatter in $\log(\mathrm{O/H})$. Assuming that $\alpha_{\min}$ completely cancels out the dependence in $\Delta\log(\mathrm{O/H})$ on SFR, $\alpha_{\min}=-\psi_1/\gamma$, where $\psi_1$ is the slope of the $\Delta\log(\mathrm{O/H)}$-SFR correlation and $\gamma$ is the MZR slope. Using $\gamma$ and $\psi_1$ values found using the O3N2 \citetalias{2004Pettini} calibration and $\mathrm{SFR_{KE12}}$ yields $\alpha_{\min}=0.3\pm0.1$, suggesting that $\alpha=0.66$ is not the optimal projection for KBSS. Neither value of $\alpha$ that can simultaneously reduce scatter in both SDSS and KBSS, suggesting that they do not occupy a common plane. 
\par Indeed, \cite{2024Garcia_a} finds that three different hydrodynamical simulations: Illustris \citep{2014Torrey}, IllustrisTNG \citep{2019Nelson}, and EAGLE \citep{2016McAlpine} all show evidence for a redshift-evolving value of $\alpha_\mathrm{min}$. Within our framework, this observation is consistent with a significant but redshift-dependent $M_\star$-$Z_g$-SFR relation \citep[which][refers to as a ``weak" FMR]{2024Garcia_a} rather than a redshift-invariant FMR. Our finding that scatter in KBSS galaxies is not reduced at the same projection as SDSS galaxies ($\alpha=0.66$) suggests that an FMR does not exist, irrespective of whether the galaxies overlap in the right panel of Figure \ref{fig:am13fmr}.
\par The analysis in Section \ref{sec:injectionrecovery} and the discussion in \cite{2024Garcia_a} both highlight the connection between parametric and non-parametric methods. An FMR should produce redshift-invariant values of $\alpha_\mathrm{min}$ and $\psi_1$ \citep[the slope of the $\Delta\log(\mathrm{O/H})$-SFR correlation][]{2024Garcia_a}. Measuring $\Delta\log(\mathrm{O/H})$ from the non-linear fit to the MZR reported in \cite{2013Andrews}, we find $\psi_1=-0.22\pm0.02$ for SDSS stacks. This slope is over four times steeper than the steepest slope found for KBSS galaxies (using \citetalias{2004Pettini} and $\mathrm{SFR_{KE12}}$, see Table \ref{tab:dmzr_dssfr}). The redshift evolution of $\psi_1$ is highly significant and shows that the strength with which SFR determines MZR scatter (see Equation \ref{res_eq}) varies with redshift.
\par Comparing KBSS to SDSS therefore demonstrates that although $z\sim0$ and $z\sim2.3$ galaxies may overlap in certain projections of $M_\star$-$Z_g$-SFR space, they are not on the same three-dimensional plane. The results of Section \ref{sec:alpha_method} show tenuous evidence for the redshift evolution of $\alpha_\mathrm{min}$, and the results of Section \ref{sec:nonparam} show strong evidence that $\psi_1$ is not redshift invariant. A redshift-invariant FMR cannot account for such evolution in $\psi_1$ implying that the $M_\star$-$Z_g$-SFR relation must evolve with redshift.
\begin{figure*}
    \centering
    \includegraphics[width=0.9\linewidth]{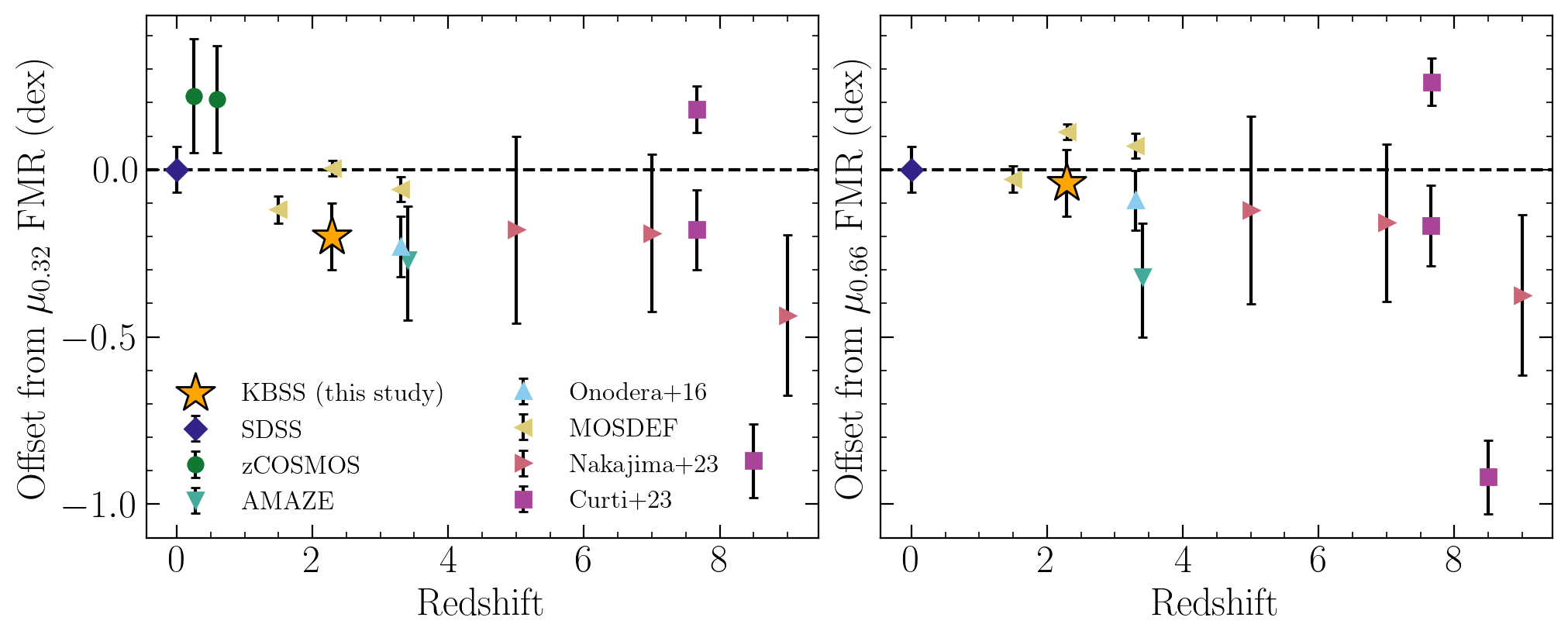}
    \caption{\emph{Left:} Offset from local galaxies at $\mu_{0.32}$ vs. redshift for the KBSS O3N2 sample (orange star), SDSS \citep[purple diamond,][]{2013Andrews}, zCOSMOS \citep[green circles,][]{2012Cresci}, AMAZE \citep[teal triangle,][]{2014Troncoso}, \citet[][light blue triangle]{2016Onodera}, MOSDEF \citep[yellow triangles,][]{2021Sanders, 2021Topping}, \citet[][pink triangles]{2023Nakajima}, and individual galaxies from \citet[][magenta squares]{2023Curti}. \emph{Right:} As in the left-hand panel, but with offset from local galaxies as $\mu_{0.66}$. While adopting $\mu_{0.66}$ bring KBSS closer to local galaxies, this is not true for all samples. Therefore, there is not a single value of $\alpha$ that minimizes scatter across redshift. }
    \label{fig:fmr_offsets}
\end{figure*}
\par In Figure \ref{fig:fmr_offsets}, we plot the mean offset between different galaxy samples and the locally defined FMR. In the left-hand panel, we define the FMR by a linear fit to SDSS galaxies in the $\mu_{0.32}$-$12+\log(\mathrm{O/H})$ plane (left panel of Figure \ref{fig:am13fmr}). We restrict the fit to galaxies with $\mu_{0.32}<10.4$ as the trend becomes non-linear at higher $\mu_{0.32}$. In the right-hand panel, we define the FMR using $\mu_{0.66}$ (right panel of Figure \ref{fig:am13fmr}) instead and restrict the linear fit to galaxies with $\mu_{0.66}<10$. 
\par As well as the KBSS galaxies analyzed in this paper, we plot the offsets of zCOSMOS \citep{2012Cresci}, MOSDEF \citep[$z\sim1.5$,][$z\sim2.3$ and $z\sim3.3$, \citealt{2021Sanders}]{2021Topping}, \cite{2016Onodera}, AMAZE \citep{2014Troncoso}, \cite{2023Nakajima}, and \cite{2023Curti}. Where possible, we use the median oxygen abundances or offsets reported in the paper, along with their associated uncertainties \citep{2012Cresci, 2014Troncoso, 2023Nakajima}. In the case of \cite{2016Onodera} and MOSDEF, we use the properties reported for composite spectra to estimate the mean offset from the FMR and take the uncertainty as the mean uncertainty on $12+\log(\mathrm{O/H})$ from the composite spectra. We plot the offset for the three individual objects with a direct-method oxygen abundance measurement \citep[SMACS0723-ID4590, 6355, and 10612 from][]{2023Curti}.
\par The galaxy samples shown here use a variety of methods to estimate oxygen abundance: some report direct-method $T_e$-based abundances \citep{2023Curti} and many of use strong-line calibrations, other than \citetalias{2004Pettini} \citep{2012Cresci, 2014Troncoso, 2016Onodera, 2018Sanders, 2023Nakajima}. We measure FMR offset from the same sample of local galaxies, but recognize that the exact value of individual FMR offsets depends on how $12+\log(\mathrm{O/H})$ and SFR are determined. When adopting $\mu_{0.32}$, KBSS galaxies are offset from the local galaxies (also see the left-hand panel of Figure \ref{fig:am13fmr}), following a trend that was previously only reported at $z\gtrsim 3.3$ \citep{2014Troncoso, 2016Onodera}. The size of the offset is comparable to the offsets measured at higher redshifts, which show systematically lower $12+\log(\mathrm{O/H})$ than predicted by the $z\sim0$ FMR. When adopting $\mu_{0.66}$, KBSS galaxies are in better agreement with local galaxies. However, this trend is not seen across all samples: for example, MOSDEF at $z\sim2.3$ and AMAZE are more offset at $\mu_{0.66}$ than at $\mu_{0.32}$. This aligns with our conclusion that although KBSS does overlap with SDSS at $\mu_{0.66}$, this is not because $\alpha_{\min}=0.66$ effectively decreases scatter at all redshifts. Indeed, Figure \ref{fig:fmr_offsets} shows that neither value of $\alpha_{\min}$ successfully minimizes scatter and FMR offset across redshift. 
\par Our results appear to be in tension with results from the MOSDEF survey \citep{2015Kriek}, which has been found to be consistent with the FMR at $z\sim1.5$ \citep{2021Topping}, $z\sim2.3$, and $z\sim3.3$ \citep{2021Sanders}. Given the similarity between the two samples, this is initially surprising. One possibility is that the different selection techniques and survey strategies have resulted in the two surveys probing distinct galaxy populations. 
\par KBSS galaxies are selected on the basis of their UV color, targeting Lyman Break analogs \citep[see][for further detail]{2004Adelberger, 2004Steidel}, and therefore more highly star-forming, and less dust-obscured galaxies are preferentially observed. The inclusion of galaxies selected according to their $\mathcal{R}$-\emph{K} color somewhat alleviates this bias \citep[][Strom et al., in prep]{2018Strom}. In contrast, MOSDEF galaxies are selected based on photo-$z$ estimates which use the Balmer break \citep{2015Kriek}, and therefore the sample may be biased towards more mature galaxies. Additionally, due to the difference in observing strategies, individual KBSS galaxies span a larger range of line ratios owing to (on average) longer exposure times. However, \cite{2021Runco} showed that the differences in the sample are generally small, and indeed when we apply our analysis, as described in Sections \ref{sec:alpha_method} and \ref{sec:nonparam}, to individual MOSDEF galaxies (see Appendix \ref{app:mosdef} for further detail), we find similar trends to those seen in KBSS galaxies: scatter is not significantly reduced at non-zero $\alpha$, and there is no significant anticorrelation between residual metallicity and residual SFR. 
\par Therefore, we conclude that the selection technique does not drive our results and instead the root of the difference likely lies in the different treatment of individual galaxies. \cite{2018Sanders}, \cite{2021Topping}, and \cite{2021Sanders} bin MOSDEF galaxies by stellar mass rather than treating galaxies individually, potentially allowing for a larger dynamic range in measured oxygen abundance. A full exploration of how binned spectra yield different results to individual galaxies is beyond the scope of this paper, but we can conclude that \emph{individual} KBSS and MOSDEF galaxies are not inconsistent with each other.
\subsection{Physical Implications}\label{sec:physimplications}
\par A redshift invariant FMR is predicted by simulations and analytical models with steady star formation, where the timescales on which the SFR and ISM enrichment fluctuate are comparable \citep{2013Dayal, 2013Lilly, 2018Torrey}. However, significant deviations from the FMR are beginning to emerge, both in high-$z$ observational data \citep[e.g.,][]{2023Curti, 2023Langeroodi, 2023Nakajima} and in hydrodynamical simulations \citep[e.g.,][]{2024Bassini, 2024Garcia_a, 2024Garcia_b}. 
\par Contrary to what is predicted by the FMR, simulations find that the strength with which SFR is correlated with MZR scatter changes with redshift. \cite{2024Garcia_a} showed that $\alpha_\mathrm{min}$ and subsequently the slope of a $\Delta\log(\mathrm{O/H})$-SFR relation (Figure \ref{fig:mzr_res}) evolve with redshift in simulations with relatively smooth SFHs. When star formation is more stochastic (``bursty"), the timescale for SFR fluctuations decreases but the timescale for ISM enrichment remains constant and independent of the SFR, further decoupling SFR and $\Delta\log(\mathrm{O/H})$ \citep{2018Torrey}. Rapid gas infall can trigger a burst of star formation and may decrease in the star formation efficiency, decoupling the correlation between gas mass and SFR, which is thought to drive the FMR. Using the Feedback in Realistic Environments simulation \citep[FIRE; ][]{2018Hopkins}, \cite{2023Bassini} found feedback to be less efficient at high-$z$, allowing galaxies to retain a high gas fraction, further decoupling SFR and ISM enrichment. 
\par As such, observed high-$z$ deviations from the FMR can be attributed to bursty star formation keeping galaxies from achieving the equilibrium state that generates the local FMR \citep[e.g.,][]{2023Curti, 2023Langeroodi, 2023Nakajima}.
The trends examined in this paper suggest that such a state of disequilibrium could persist to $z\sim2.3$, explaining the deviation of KBSS galaxies from the $z\sim0$ FMR (Figure \ref{fig:am13fmr}) and the very weak correlation between residual metallicity and residual SFR (Figures \ref{fig:mzr_res} and \ref{fig:res_res}).

\par FIRE simulations have also suggested that the MZR's evolution from $z\sim0-3$ is not driven by changes to the gas fraction (observationally probed by SFR), but is instead driven by the enrichment of inflows and outflows, relative to the enrichment of the ISM \citep{2024Bassini}. This suggests that SFR (and by proxy, gas fraction) is not the most fundamental third parameter. The weak correlation between residual metallicity and residual sSFR in KBSS galaxies could be consistent with idea that gas fraction may not be the primary driver of galaxies' chemical evolution. Alternatively, a lower star formation efficiency due to rapid gas infall, this could also weaken the relationship between SFR and gas fraction, resulting in a weaker correlation between residual metallicity and residual sSFR. At present, we lack the data needed to break the degeneracy between these two cases. Further observations would be needed in order to either measure the metallicity of infalling and outflowing gas or to measure the gas fraction. 


\section{Conclusions}\label{sec:conclusions}
\par We have used both parametric (Section \ref{sec:alpha_method}) and non-parametric (Section \ref{sec:nonparam}) methods to investigate the relationship between stellar mass, gas-phase oxygen abundance, and SFR and its consistency with the FMR in 273 galaxies from KBSS-MOSFIRE. We use strong-line calibrations to estimate gas-phase oxygen abundance and test eight different calibrations (each of which produces an MZR with a different shape and normalization, see Section \ref{sec:mzr} and Figure \ref{fig:o3n2_mzrs}) to mitigate the potential bias introduced by a single calibration. 
To account for the effect of subsolar stellar metallicities on the ionizing stellar output, we adopt a metallicity-dependent conversion factor betweem H$\alpha$ luminosity and SFR (Section \ref{sec:sfrs}). For comparison, we also conduct our analysis using SFRs calculated using a locally-calibrated constant conversion factor.
\par The results of our analysis lead us to conclude that
\begin{enumerate}
	\item A metallicity-dependent SFR conversion factor results in a lower SFR for more metal-poor galaxies than would be predicted using a constant conversion factor (Figure \ref{fig:sfr_hist}). The associated SFMS may be steeper and offset towards lower SFR (Figure \ref{fig:sfms_hists}). The size of this effect is contingent on the strong-line calibration used to estimate metallicity: calibrations associated with a steeper MZR produce a slightly steeper SFMS (Section \ref{sec:sfrs}).
    \item Across eight strong-line calibrations and two SFR calculations, the parametric method suggests that KBSS galaxies are inconsistent with the FMR (Section \ref{sec:alpha_method}). Adopting a non-zero value of $\alpha$ does not result in a significant reduction in scatter (Figure \ref{fig:o3n2_alpha}). 
	\item Across eight strong-line calibrations calculations, KBSS galaxies show weak and marginally significant anticorrelations both between residual metallicity and absolute SFR$_\mathrm{KE12}$ (top panel, Figure \ref{fig:mzr_res}) and between residual metallicity and residual sSFR$_\mathrm{KE12}$ (top panel, Figure \ref{fig:res_res}). Adopting SFR$(Z)$ results in a weaker and less significant anticorrelation (bottom panels of Figures \ref{fig:mzr_res} and \ref{fig:res_res}, and Section \ref{sec:nonparam}), suggesting there may not be a significant $M_\star$-SFR$(Z)$-$Z_g$ correlation in the KBSS sample.
	\item Injection-recovery tests based on the demographics of our galaxy sample show that the parametric and non-parametric approaches are only concordant if the anticorrelation between residual metallicity and SFR is significantly stronger than what is observed at $z\sim2.3$ (Figure \ref{fig:slope_scatter_grid} and Section \ref{sec:injectionrecovery}).
    \item KBSS galaxies are inconsistent with the locally-defined FMR (Sections \ref{sec:am13_comp} and \ref{sec:othersamples}). The significant redshift evolution of the $\Delta\log(\mathrm{O/H})$-SFR anticorrelation (Section \ref{sec:othersamples}), lack of scatter reduction at any $\alpha\neq0$ (Figure \ref{fig:o3n2_alpha}), and offset towards lower gas-phase oxygen abundance at fixed $\mu_{0.32}$ (Figure \ref{fig:am13fmr}) all point towards an evolving $M_\star$-$Z_g$-SFR relationship.
	\item By comparing to results from hydrodynamical simulations, our results suggest that KBSS galaxies may not be in the equilibrium state of the gas regulator model, and that the disequilibrium state that has already been observed at higher redshifts persists to $z\sim2.3$ (Section \ref{sec:physimplications}). 
	\item The weak anticorrelation between residual metallicity and residual sSFR is consistent with two possible scenarios: rapid gas infall has caused gas fraction and SFR to become less tightly coupled, or gas fraction is not the most fundamental third parameter driving the evolution of the MZR (Section \ref{sec:physimplications}). 
\end{enumerate}
\par This work highlights the need to be consistent and precise when discussing the FMR as different conclusions can arise depending on the choice of method. Further work is needed in order to understand the nature of the baryon cycle at cosmic noon and why it might differ from the equilibrium conditions that give rise to the FMR. The results presented here are not in tension with alternative explanations of the MZR's evolution but further work is needed in order to determine whether this is the case and identify other potential third parameters.
\section*{Acknowledgments}
This work is based on data obtained at the W. M. Keck Observatory, operated as a scientific partnership among the California Institute of Technology, the University of California, and the National Aeronautics and Space Administration. This work made use of v2.2.1 of the Binary Population and Spectral Synthesis (BPASS) models as described in \cite{2017Eldridge} and \cite{2018Stanway}.
The Observatory was made possible by the generous financial support of the W. M. Keck Foundation. T.B.M. was supported by a CIERA Fellowship. N.K.C. would like to thank Claude-Andr\'e Faucher-Gigu\`ere and Andrew Marszewski for their insightful comments regarding results from the FIRE simulations, as well as Adam Carnall for his insight on the Bagpipes SED-fitting code. Lastly, the authors wish to recognize and acknowledge the very significant cultural role and reverence that the summit of Maunakea has always had within the Native Hawaiian community. We are most fortunate to have the opportunity to conduct observations from this mountain. 
%





\appendix
\section{N2 Calibrations}\label{app:n2}
\begin{figure*}
    \centering
    \includegraphics[width=0.9\linewidth]{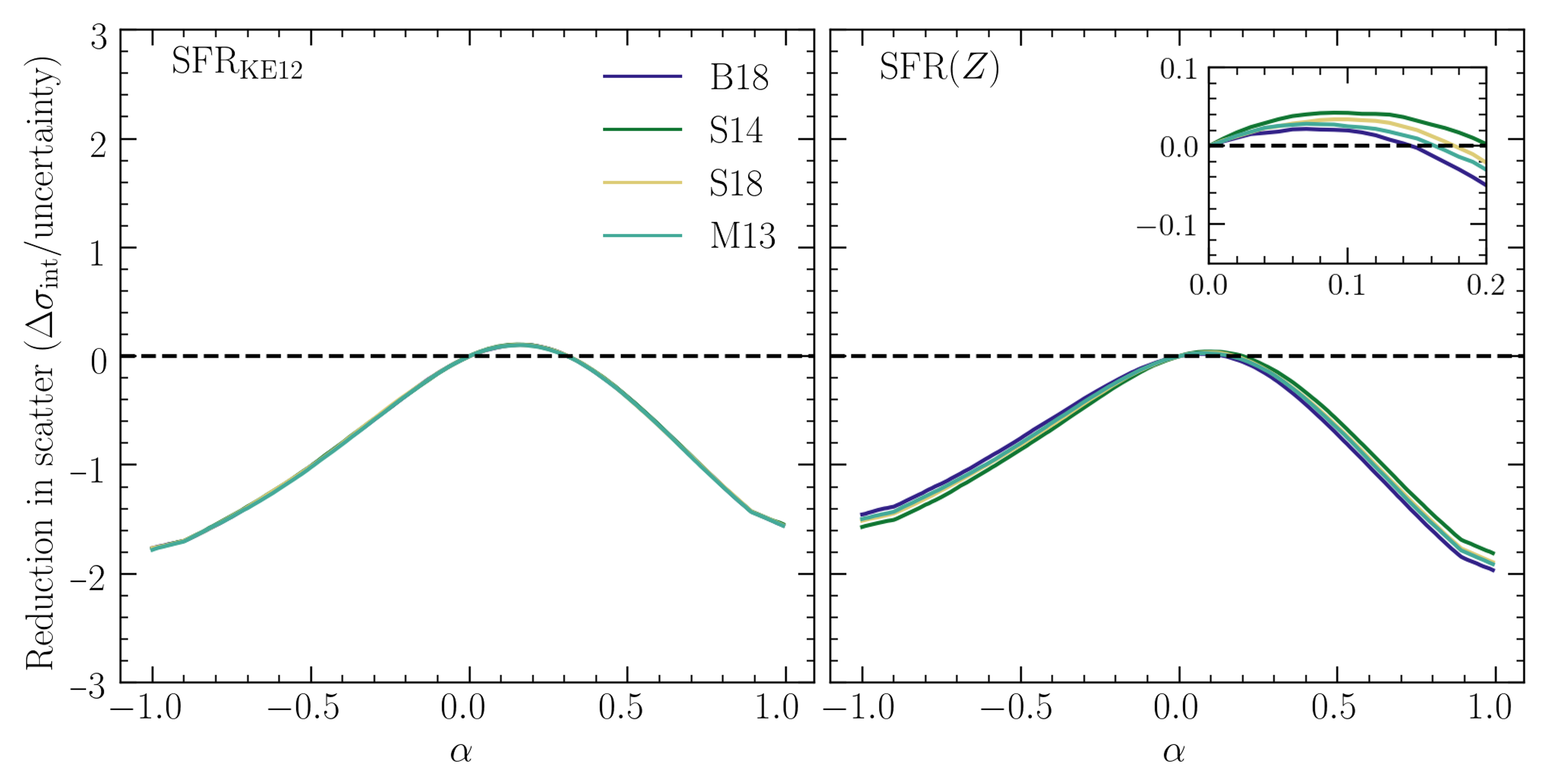}
    \caption{Reduction in scatter vs. $\alpha$ for KBSS galaxies, as in Figure \ref{fig:o3n2_alpha}, but using N2 calibrations. Each solid curve shows a different calibration, the left-hand panel uses $\mathrm{SFR_{KE12}}$, and the right-hand panel uses $\mathrm{SFR}(Z)$. The inset panel shows the same data as the right-hand panel but limited to $0\leq\alpha\leq0.2$ in order to show the effect of adopting SFR$(Z)$. As was found using O3N2 calibrations, scatter is not significantly reduced at $\alpha\neq0$.}
    \label{fig:n2_alpha}
\end{figure*}
\begin{table*}[]
    \centering
    \caption{The slope ($\psi_1$, $\psi_2$), Spearman's rank correlation coefficient ($\rho_1$, $\rho_2$), $p$-value ($p_1$, $p_2$) for the $\Delta\mathrm{MZR}_{\mathrm{N2}}$-SFR and $\Delta\mathrm{MZR}_{\mathrm{N2}}$-$\Delta$sSFR anticorrelations respectively.}
    \begin{tabular}{c c| c c c| c c c}
        \hline\hline
        \multirow{2}{*}{Calibration} & \multirow{2}{*}{SFR} & \multicolumn{3}{c|}{$\Delta\log(\mathrm{O/H})$-SFR} & \multicolumn{3}{c}{$\Delta\log(\mathrm{O/H})$-$\Delta$sSFR} \\ 
         &                      & $\psi_1$    & $\rho_1$    & $p_1$   & $\psi_2$       & $\rho_2$       & $p_2$      \\\hline\hline
         \citetalias{2013Marino} & \citetalias{2012Kennicutt} & $-0.03\pm0.02$ & $-0.08$ & 0.2 & $-0.02\pm0.02$ & $-0.09$ & 0.1\\
         \citetalias{2013Marino} & Z & $-0.02\pm0.02$& $-0.04$ & 0.5 &$-0.01\pm0.02$ & $-0.05$ & 0.4\\
         \citetalias{2014Steidel} & \citetalias{2012Kennicutt} & $-0.02\pm0.01$& $-0.09$ & 0.2 & $-0.02\pm0.01$ & $-0.09$ & 0.1\\
        \citetalias{2014Steidel} & Z & $-0.02\pm0.01$& $-0.06$ & 0.4 & $-0.01\pm0.01$ & $-0.06$ & 0.3\\
        \citetalias{2018Bian} & \citetalias{2012Kennicutt} & $-0.03\pm0.02$ & $-0.08$ & 0.2 &$-0.03\pm0.02$ & $-0.09$ & 0.1\\
        \citetalias{2018Bian} & Z & $-0.03\pm0.02$& $-0.03$ & 0.7 & $-0.009\pm0.02$ & $-0.04$ & 0.5\\
        \citetalias{2018Strom} & \citetalias{2012Kennicutt} & $-0.02\pm0.01$& $-0.08$ & 0.2 & $-0.02\pm0.01$ & $-0.09$ & 0.1\\
        \citetalias{2018Strom} & Z &$-0.01\pm0.01$ & $-0.04$ & 0.5 & $-0.008\pm0.01$ & $-0.05$ & 0.4\\
        \hline\hline
    \end{tabular}
    \label{tab:n2_res}
\end{table*}
\begin{figure}
    \centering
    \includegraphics[width=1\linewidth]{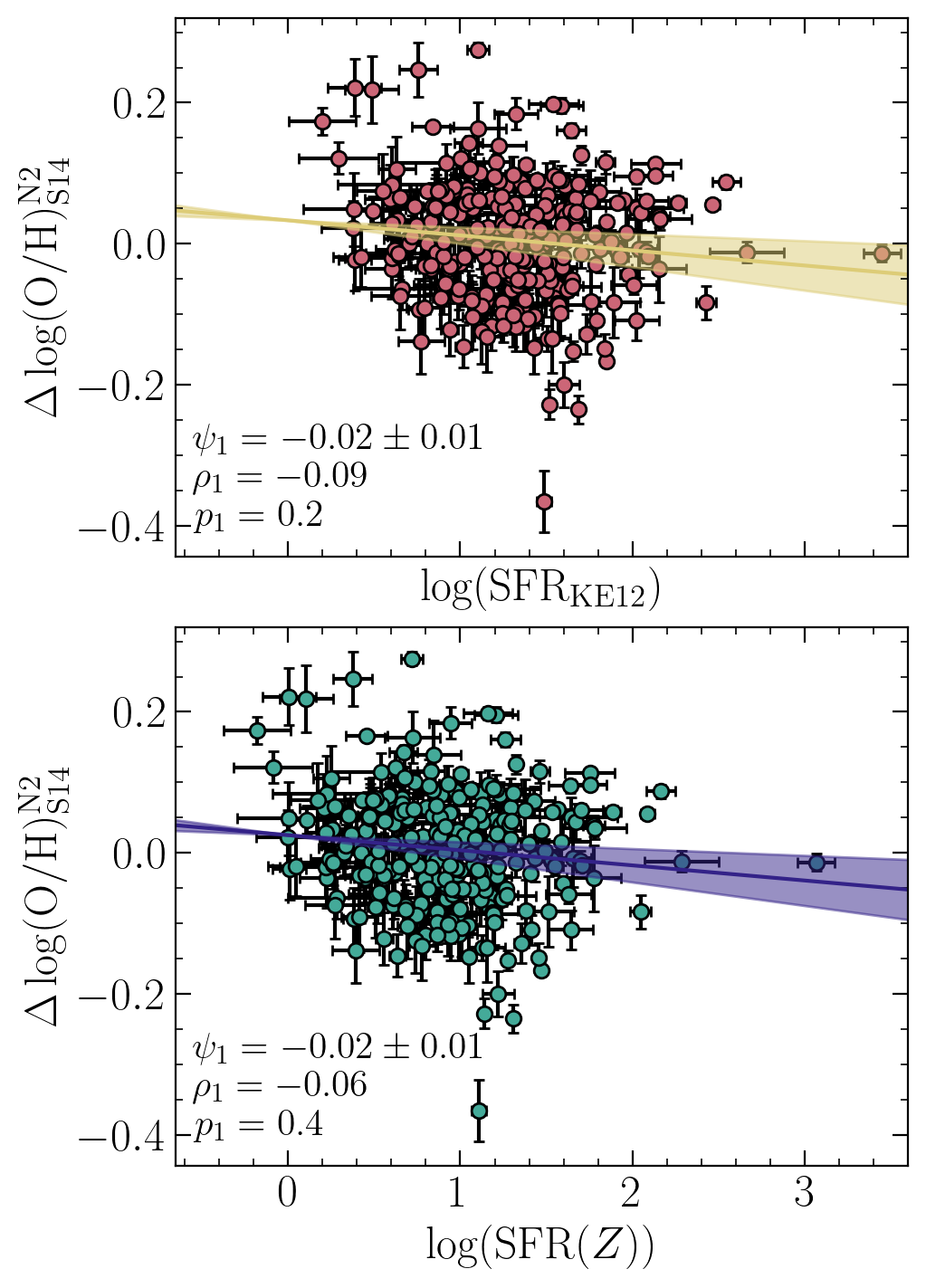}
    \caption{$\Delta\log(\mathrm{O/H})$ vs. $\log(\mathrm{SFR})$ (top: $\mathrm{SFR_{KE12}}$, bottom: $\mathrm{SFR}(Z)$) for KBSS galaxies, as in Figure \ref{fig:mzr_res} but using the \citetalias{2014Steidel} N2 calibration. There no significant anticorrelation between $\Delta\mathrm{MZR_{N2}}$ and $\log(\mathrm{SFR})$.}
    \label{fig:n2_sfrres}
\end{figure}

\begin{figure}
    \centering
    \includegraphics[width=1\linewidth]{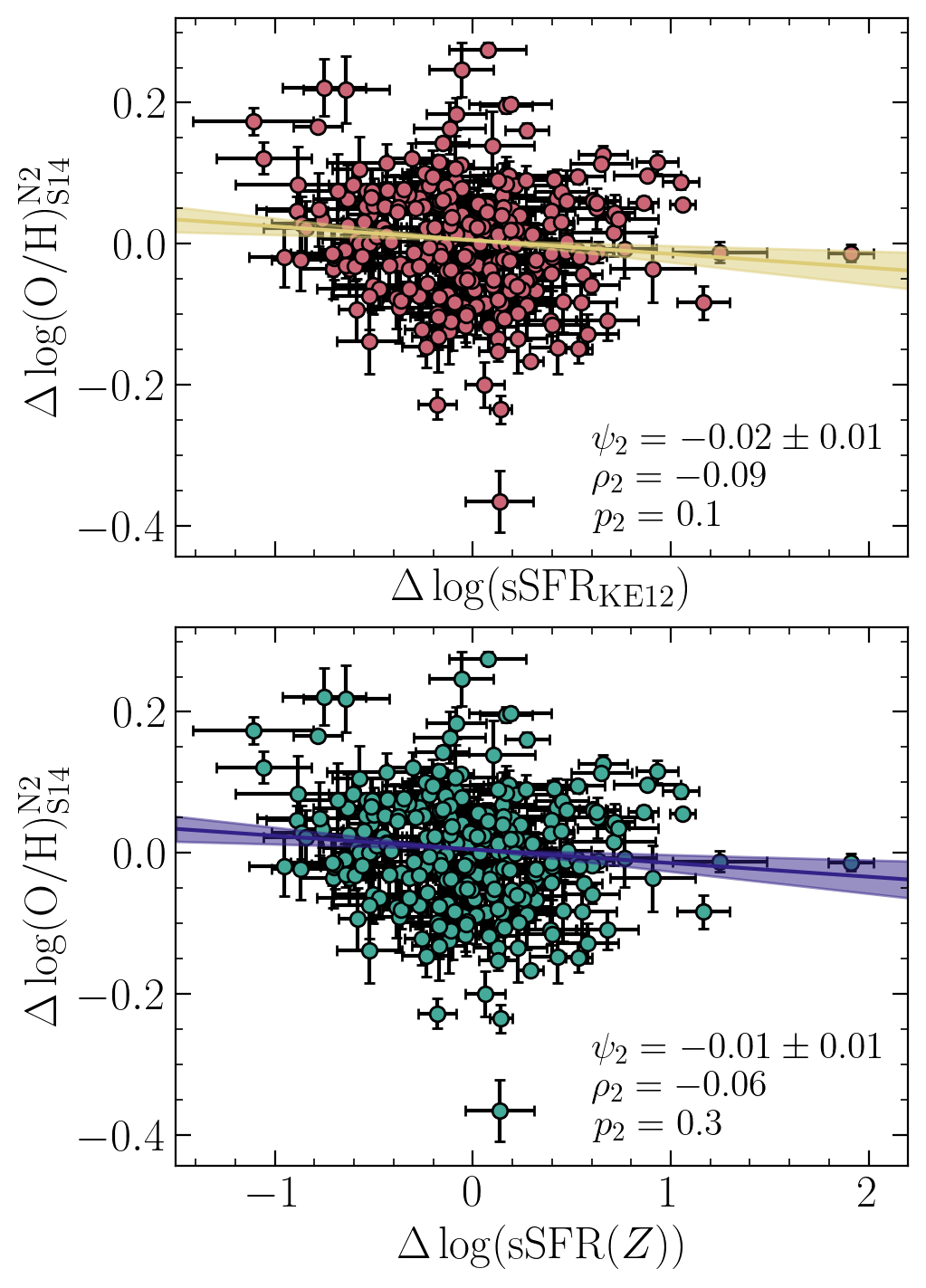}
    \caption{$\Delta\log(\mathrm{O/H})$ vs. $\Delta(\mathrm{sSFR})$ (top: $\mathrm{SFR_{KE12}}$, bottom: $\mathrm{SFR}(Z)$) for KBSS galaxies, as in Figure \ref{fig:res_res} but using the \citetalias{2014Steidel} N2 calibration.
    There is no significant anticorrelation between $\Delta\mathrm{MZR_{N2}}$ and $\Delta\log(\mathrm{sSFR})$.}
    \label{fig:n2_resres}
\end{figure}

\par High-$z$ galaxies are offset from local galaxies in the N2-BPT due to an increase in spectral hardness at fixed gas-phase oxygen abundance \citep{2014Masters, 2014Steidel, 2015Shapley, 2017Strom, 2024Shapley_a}, showing that changes in N2 are sensitive to the spectral hardness, not only gas-phase oxygen abundance. O3N2 is sensitive to ionisation parameter \citep{2002Kewley}, but in KBSS, O3N2 has been found to be a less biased indicator of oxygen abundance than N2 \citep{2014Steidel}, motivating our preference for O3N2 in the main body of this paper. However, N2 requires the observation of just two emission lines with a small wavelength separation, removing the need for multi-band observations. As a result, N2 is frequently used throughout the literature. Hence, while we do not present N2-based results in the main body of the paper, we include them here for completeness.

\par Figure \ref{fig:n2_alpha} shows the results of the parametric analysis detailed in Section \ref{sec:alpha_method}. As in Figure \ref{fig:o3n2_alpha}, we plot reduction in scatter, relative to the MZR scatter, against $\alpha$. Scatter is not significantly minimized by any $\alpha\neq0$, suggesting that the N2 sample is inconsistent with the FMR. In fact, the reduction in scatter in the N2 sample is smaller than in the O3N2 sample.

\par Figures \ref{fig:n2_sfrres} and \ref{fig:n2_resres} show the results of the non-parametric analysis detailed in Section \ref{sec:nonparam}, when adopting the \citetalias{2014Steidel} calibration. Across calibrations and SFR estimates, we do not find significant $\Delta\log(\mathrm{O/H})$-SFR and $\Delta\log(\mathrm{O/H})$-$\Delta$sSFR (Table \ref{tab:n2_res}) anticorrelations. N2-based results can therefore be used to draw the same conclusions as the O3N2 sample: KBSS galaxies are inconsistent with the locally-defined FMR and may show a weak $M_\star$-$Z_g$-SFR relation. 

\section{Comparison to MOSDEF}\label{app:mosdef}
\par MOSDEF \citep{2015Kriek} is the most comparable survey to KBSS, targetting the rest-optical spectra of $z\sim2$ galaxies using the MOSFIRE spectrograph. As a result, it is useful to compare the results inferred from these two different surveys. As discussed in Section \ref{sec:othersamples}, the surveys are based on different selection methods but are generally very similar in terms of stellar mass, dust extinction, and galaxy properties \citep{2021Runco}. 
\begin{figure*}
    \centering
    \includegraphics[width=1.0\linewidth]{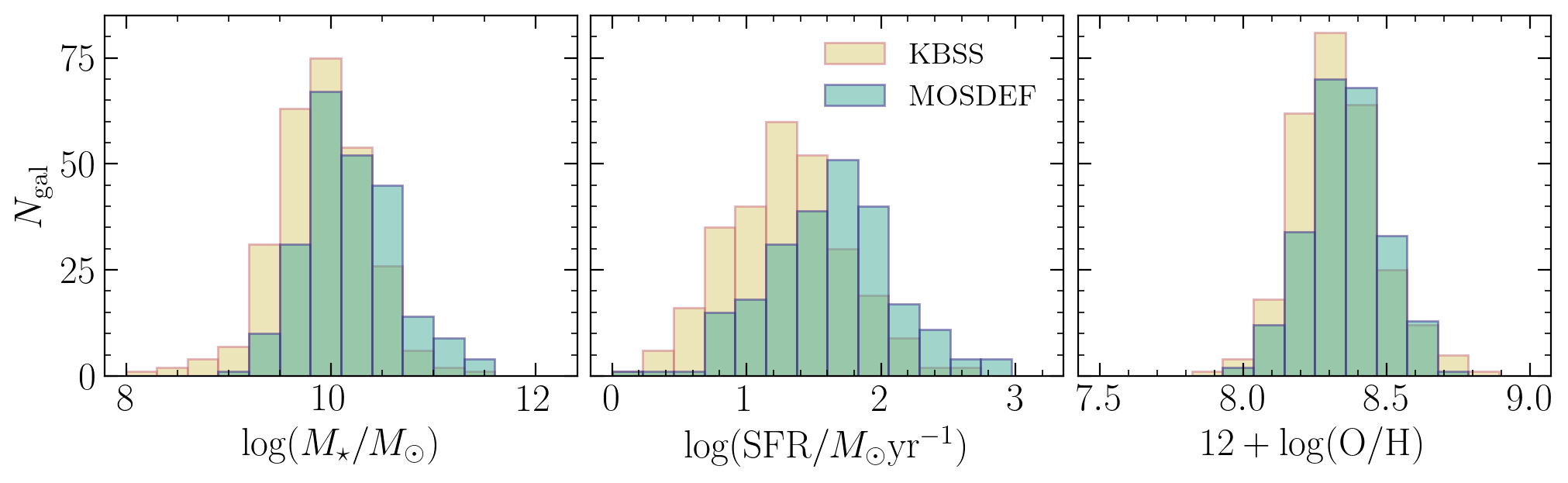}
    \caption{Histograms showing the distribution of (left to right) stellar mass, SFR, and oxygen abundance in KBSS (yellow) and MOSDEF (teal). MOSDEF includes more high-mass, more highly star forming galaxies than KBSS, although the distribution of $12+\log$(O/H) is comparable between the two samples. The $M_\star$-$Z_g$-SFR correlation may be stronger for higher mass galaxies. }
    \label{fig:mosdef_hist}
\end{figure*}
\par Despite these similarities, our conclusion that KBSS galaxies do not follow the FMR is in tension with reported MOSDEF results \citep{2021Sanders, 2021Topping}. The tension appears to originate in the different treatments of individual galaxy spectra: previous MOSDEF results are reported based on stacked spectra, while we use individual galaxies. Stacked spectra may allow for a larger dynamic range in $\log(\mathrm{O/H})$, particularly at low metallicity, since stacking allows for weaker emission lines to be measured at a higher signal to noise. However, stacking makes it difficult to measure intrinsic scatter within a galaxy population. Due to the difference in observing strategies, individual KBSS spectra span a larger dynamic range in line ratios than individual MOSDEF spectra. As a result, using individual galaxies instead of stacks may more drastically curtail the dynamic range of $\log(\mathrm{O/H})$ in MOSDEF than in KBSS, potentially explaining why MOSDEF results are significantly different when individual galaxies are analyzed. Here, we show that when we apply the analysis outlined in Sections \ref{sec:alpha_method} and \ref{sec:nonparam} to individual MOSDEF galaxies, the results are in fact aligned. 
\par Using the \href{https://mosdef.astro.berkeley.edu/for-scientists/data-releases/}{line measurements} made public by the MOSDEF team, we impose the same SNR cuts ($\mathrm{SNR}>3$ for H$\alpha$, H$\beta$, [O \textsc{iii}]$\lambda5007$, and [N \textsc{ii}]$\lambda 6885$, and $\mathrm{SNR}>5$ for $\mathrm{H}\alpha/\mathrm{H}\beta$). We take the stellar masses from the 3D-HST catalog \citep{2014Skelton}, which also assume a \cite{2003Chabrier} IMF, and make a further cut to exclude galaxies with $\log(M_\star/M_\odot)<9$ as the sample is incomplete below this mass \citep{2021Topping}. To ensure we construct a comparable sample to the KBSS sample presented in this paper, we also restrict the sample to galaxies in the redshift range $1.90\leq z\leq2.61$. After making these cuts, we obtained a sample of 233 galaxies with well-measured O3N2 and 260 galaxies with well-measured N2. The distribution in stellar mass, SFR, and oxygen abundance in the O3N2 sample can be seen in Figure \ref{fig:mosdef_hist}. Since no characteristic uncertainty is reported for these stellar masses, we adopt a conservative estimate of 0.1 dex as the uncertainty on stellar mass, equal to the median uncertainty for KBSS stellar masses. In reality, we recognize that the true uncertainty is likely lower, due to the more extensive photometry available for the stellar mass estimates. SFR is estimated from the H$\alpha$ luminosity, using both the \citetalias{2012Kennicutt} conversion factor and a metallicity-dependent conversion factor. 
\par \cite{2021Runco} found the spectroscopic KBSS and MOSDEF samples to be comparable, with some slight differences in the median stellar mass. Once our quality cuts are made, this difference becomes more pronounced, and we also see a significant difference in the median SFR (see Figure \ref{fig:mosdef_hist}). Compared to the KBSS O3N2 sample, the MOSDEF O3N2 sample has a higher median stellar mass ($1.35\times10^{10}M_\odot$, 0.3 dex larger than the median mass in the KBSS) and a higher SFR ($44\, M_\odot \mathrm{yr}^{-1}$, 0.15 dex higher than in KBSS). However, a full exploration of the effect of sample differences on the nature of the $M_\star$-$Z_g$-SFR relationship is beyond the scope of this paper.
\begin{figure*}
    \centering
    \includegraphics[width=1.0\linewidth]{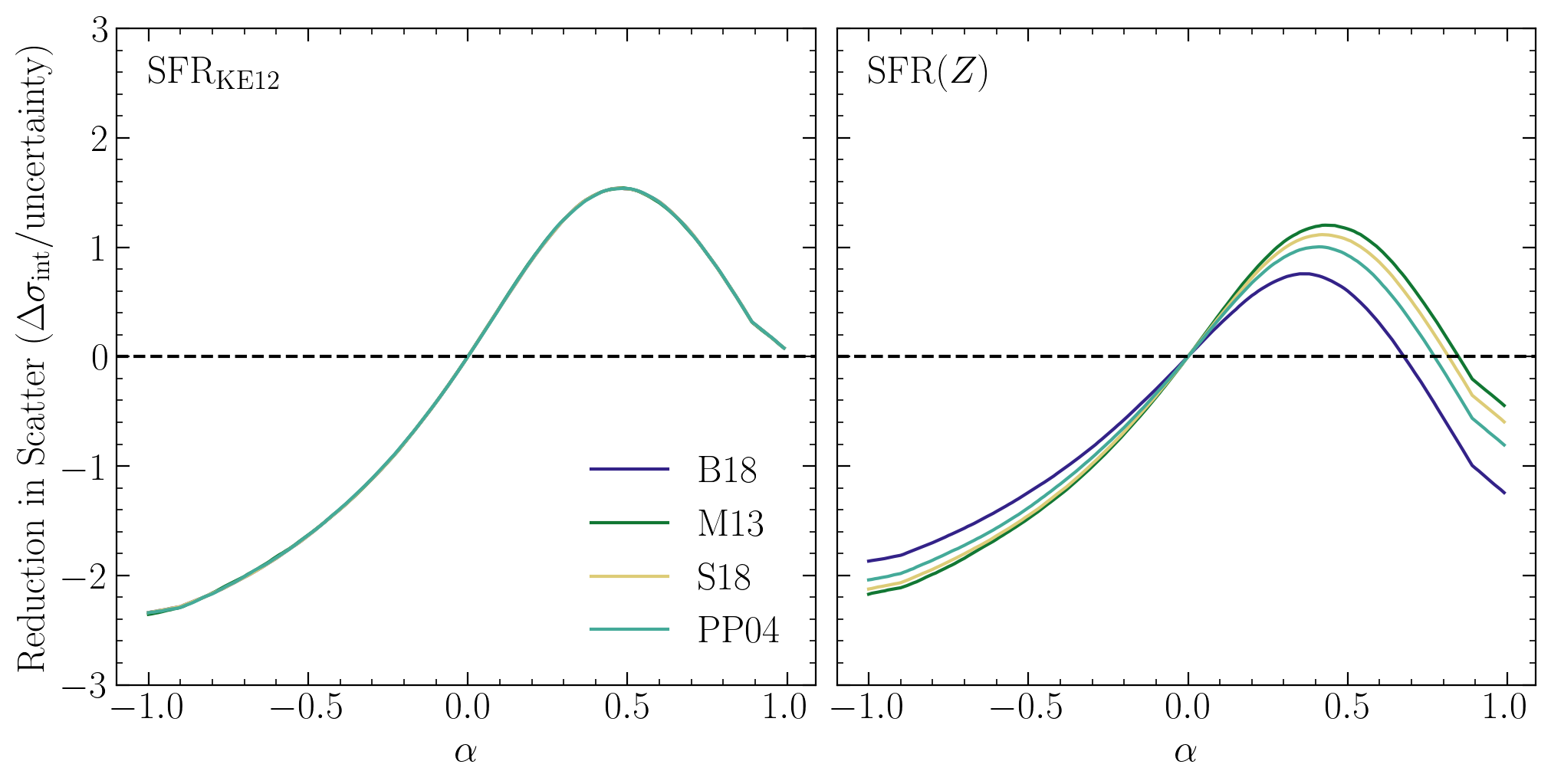}
    \caption{Reduction in scatter vs. $\alpha$, as in Figures \ref{fig:o3n2_alpha} and \ref{fig:n2_alpha}, but for MOSDEF galaxies, using $\mathrm{SFR_{KE12}}$ (left) and $\mathrm{SFR}(Z)$ (right). Each solid curve shows a different O3N2 calibration. As in the KBSS sample, scatter is not significantly reduced by any value of $\alpha$, and scatter reduction is suppressed by adopting $\mathrm{SFR}(Z)$.}
    \label{fig:mosdef_alpha}
\end{figure*}
\begin{figure}
    \centering
    \includegraphics[width=1.0\linewidth]{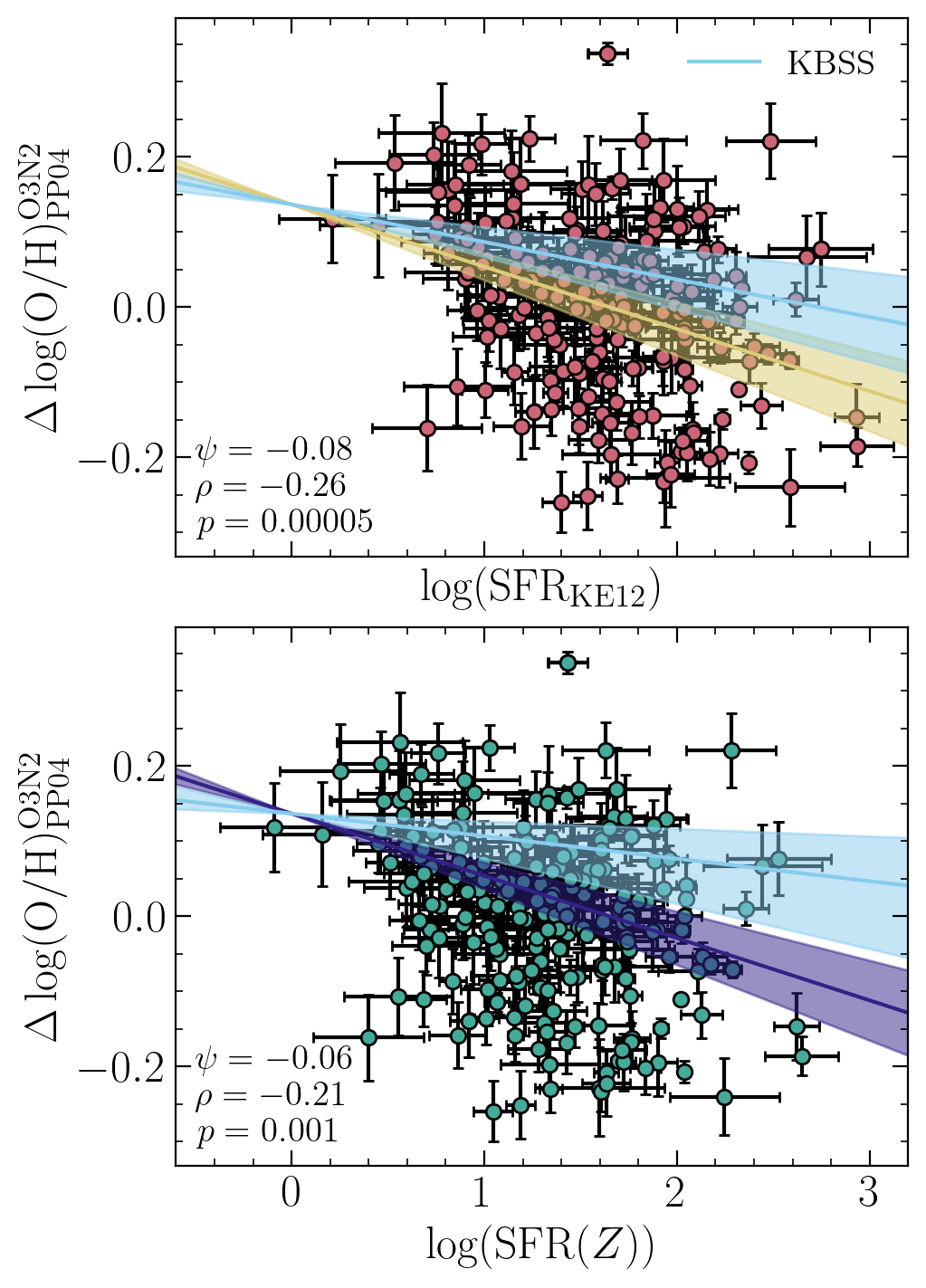}
    \caption{$\Delta\log(\mathrm{O/H})$ vs. $\log(\mathrm{SFR_{KE12}})$, as in Figures \ref{fig:mzr_res} and \ref{fig:n2_sfrres}, but using MOSDEF galaxies. The top panel uses $\mathrm{SFR_{KE12}}$ and the bottom panel uses SFR$(Z)$. The slopes found for KBSS (Figure \ref{fig:mzr_res}) are shown in light blue. The anticorrelation is stronger and more significant than what is found for KBSS but still shallower than what is found for local galaxies, implying a redshift evolution.}
    \label{fig:res_sfr_mosdef}
\end{figure}
\begin{figure}
    \centering
    \includegraphics[width=1.0\linewidth]{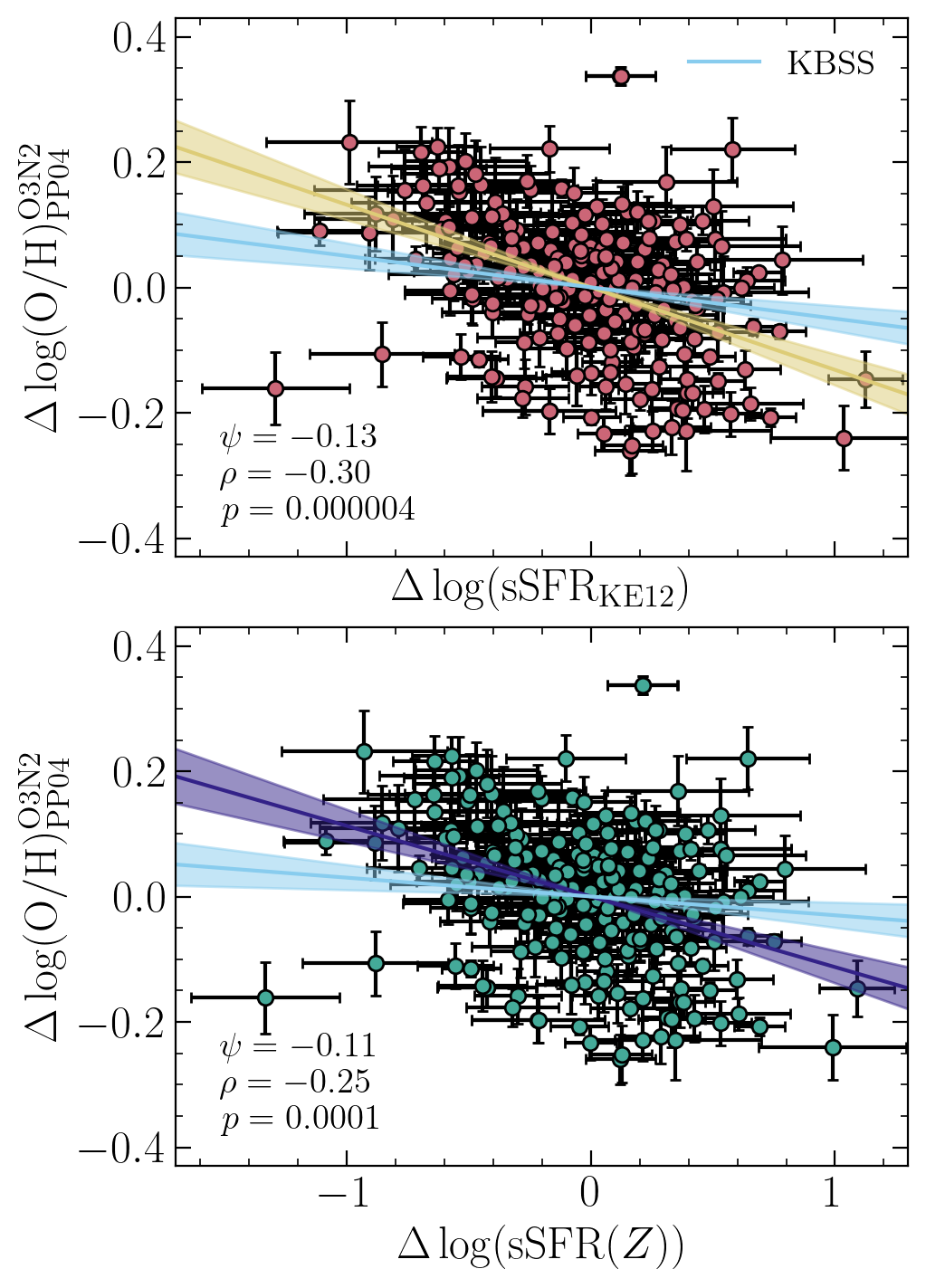}
    \caption{$\Delta\log(\mathrm{O/H})$ vs. $\Delta\mathrm{sSFR_{KE12}}$, as in Figures \ref{fig:res_res} and \ref{fig:n2_resres}, but using MOSDEF galaxies. The top panel uses $\mathrm{SFR_{KE12}}$ and the bottom panel uses SFR$(Z)$. For comparison, the KBSS regression line (Figure \ref{fig:res_res}) is shown in light blue. As in Figure \ref{fig:res_sfr_mosdef}, the anticorrelation is stronger and more significant than in KBSS and weaker than in local galaxies, implying a redshift evolution.}
    \label{fig:eres_res_mosdef}
\end{figure}
\par First, we apply the parametric method, which measures the scatter in the correlation between $12+\log(\mathrm{O/H})$ and $\mu_\alpha$. See Section \ref{sec:alpha_method} for further detail regarding the parametric method. Figure \ref{fig:mosdef_alpha} shows the reduction in scatter at different values of $\alpha$. When using $\mathrm{SFR_{KE12}}$ (cf. left panel of Figure \ref{fig:mosdef_alpha}), scatter is minimised by 1.5$\sigma$ at $\alpha\sim0.5$. This level of reduction is larger than in KBSS (left panel of Figure \ref{fig:o3n2_alpha}), although still only marginally significant. As is seen in KBSS (right panel of Figure \ref{fig:o3n2_alpha}) and explained in Section \ref{sec:nonparam}, adopting SFR$(Z)$ suppresses the reduction in scatter. Strong-line calibrations with a larger $|B|$ factor (see Table \ref{tab:calibrations}) are associated with a stronger suppression of the reduction in scatter. In the case of \citetalias{2004Pettini} and \citetalias{2018Bian}, scatter is reduced by $<1\sigma$, and in the case of \citetalias{2013Marino} and \citetalias{2018Strom}, the reduction in scatter remains marginal at $<1.5\sigma$. Thus, while MOSDEF galaxies do show a more significant reduction in scatter than KBSS galaxies, the results remain consistent and we cannot confidently conclude that MOSDEF galaxies show a significant reduction in scatter when incorporating SFR as a third parameter.
\par Next, we use the non-parametric method (Section \ref{sec:nonparam}) to quantify any $\Delta\log(\mathrm{O/H})$-SFR and $\Delta\log(\mathrm{O/H})$-$\Delta$sSFR correlations. More significant differences between MOSDEF and KBSS emerge here. Across strong-line calibrations (both O3N2- and N2-based) and SFR calculations, we measure significant $\Delta\log(\mathrm{O/H})$-SFR and $\Delta\log(\mathrm{O/H})$-$\Delta$sSFR anticorrelations. Given the larger reduction in scatter seen in MOSDEF (Figure \ref{fig:mosdef_alpha}) and the analysis presented in Section \ref{sec:injectionrecovery}, this is unsurprising. MOSDEF galaxies are associated with steeper $\Delta\log(\mathrm{O/H})$-SFR (Figure \ref{fig:res_sfr_mosdef}) and $\Delta\log(\mathrm{O/H})$-$\Delta$sSFR (Figure \ref{fig:eres_res_mosdef}) slopes, more negative Spearman-$\rho$ values, and a higher statistical significance. Moreover, we recover a similar $\Delta\log(\mathrm{O/H})$-$\Delta$sSFR slope to \cite{2018Sanders}, who found a slope of $-0.14\pm 0.034$ for $z\sim2.3$ MOSDEF galaxies. 
\par Notably, the $\Delta\log(\mathrm{O/H})$-$\Delta$sSFR slope is much shallower than the slope associated with $z\sim0$ SDSS stacks, as is discussed in Section \ref{sec:othersamples}, this implies that MOSDEF galaxies are also inconsistent with a redshift-invariant FMR. The fact that MOSDEF galaxies in our sample and SDSS galaxies generally have higher stellar masses and SFRs than KBSS galaxies may suggest that $M_\star$-$Z_g$-SFR correlations are stronger in higher mass galaxies. 


\bibliography{sample631}{}
\bibliographystyle{aasjournal}



\end{document}